\PassOptionsToPackage{dvipsnames}{xcolor} 
\documentclass{fmcad}

\usepackage[dvipsnames]{xcolor}
\usepackage{amsmath, amssymb, amsthm}   
\usepackage{graphics}
\usepackage{wrapfig}
\usepackage{caption}
\usepackage{verbatim}
\usepackage{paralist}
\usepackage{cancel}
\usepackage{xspace}
\usepackage{multicol}

\definecolor{navy}{RGB}{0,0,128}

\newcommand{\relu}{\text{ReLU}\xspace{}}
\newcommand{\coef}{\text{\emph{coef}}\xspace{}}

\newcommand{\basic}{\mathcal{B}}
\newcommand{\allvars}{\mathcal{X}}
\newcommand{\ub}{u}
\newcommand{\lb}{l}
\newcommand{\farkasUpper} [1] {f_u(#1)}
\newcommand{\farkasLower} [1] {f_l(#1)}

\newcommand{\assignment}{\alpha{}}

\usepackage{tikz,ifthen,pgfplots}
\usetikzlibrary{arrows,trees,backgrounds,automata,shapes,decorations,plotmarks,fit,calc,positioning,shadows,chains}
\tikzstyle{every pin edge}=[<-,shorten <=1pt]
\tikzstyle{neuron}=[circle,fill=black!25,minimum size=17pt,inner sep=0pt]
\tikzstyle{input neuron}=[neuron, fill=green!50]
\tikzstyle{output neuron}=[neuron, fill=red!50]
\tikzstyle{hidden neuron}=[neuron, fill=blue!50]
\tikzstyle{merged neuron}=[neuron, fill=orange!50]
\tikzstyle{annot} = [text width=6em, text centered]
\tikzstyle{nnedge} = [-{stealth},shorten >=0.1cm, shorten <=0.05cm,line width=0.8pt,black]
\tikzstyle{proofNode} = [rounded rectangle, fill=purple!30]
\tikzstyle{proofEdge} = [-{stealth},shorten >=0.1cm, shorten <=0.05cm,line width=0.8pt,black]
\usetikzlibrary{calc}

\newcommand{\rulename}[1]{\ensuremath{\mathsf{#1}}\xspace}
\newcommand{\irulename}[2]{\ensuremath{\mathsf{#1}_{#2}}\xspace}
\newcommand{\pivot}[1]{\irulename{Pivot}{#1}}
\newcommand{\failureSlack}{\rulename{Failure_1}}
\newcommand{\failureBounds}{\rulename{Failure_2}}
\newcommand{\update}{\rulename{Update}}

\newcommand{\success}{\rulename{Success}}

\newcommand{\slackPlus}{\text{slack}^+\xspace{}}
\newcommand{\slackMinus}{\text{slack}^-\xspace{}}
\newcommand{\slack}{\text{slack}\xspace{}}

\newcommand{\pivotOperation}{\textit{pivot}}
\newcommand{\updateOperation}{\textit{update}}

\newcommand{\mysubsection}[1]{\medskip\noindent\textbf{#1}}

\newcommand{\drule}[2]{
	\renewcommand{\arraystretch}{1.2}
	\(\begin{array}{c}
		#1 \\
		\hline 
		#2
	\end{array}\)
}

\newcommand{\sat}{\texttt{SAT}}
\newcommand{\unsat}{\texttt{UNSAT}}

\newcommand{\rn}[1]{\mathbb{R}^{#1}}
\newcommand{\sumallvars}{\underset{j \neq i }{\overset{ }{\sum} } c_j \cdot x_j}
\newcommand{\sumallnbvars}{\underset{j \notin \basic }{\overset{ }{\sum} } c_j \cdot x_j}
\newcommand{\sumposvars}[1]{\underset{j \neq i, c_j > 0 }{\overset{ }{\sum} } c_j \cdot {#1}}
\newcommand{\sumnegvars}[1]{\underset{j \neq i, c_j < 0 }{\overset{ }{\sum} } c_j \cdot {#1}}

\newcommand{\mainthm}{  
	Let $ A\cdot V = 0 $ such that $ l \leq V \leq u $ be an LP instance,
	where $ A \in M_{m \times n} (\mathbb{R})$  and $ l,V,u \in \rn{n} $.
	
	Let $u', l'\in \rn{n} $ represent
	\textit{dynamically tightened} bounds of $ V $.  Then
	$ \forall i \in \left[ n\right] \: \exists \farkasUpper{x_i}, \farkasLower{x_i} \in
	\rn{m}$ such that
	$ \farkasUpper{x_i}^\intercal \cdot A$ and
        $\farkasLower{x_i}^\intercal \cdot A$
	can be used to efficiently compute $ u'(x_i), l'(x_i) $
        from $l$ and $u$.  Moreover, 
	vectors $\farkasUpper{x_i}$ and $\farkasLower{x_i}$ can be constructed during the run of the Simplex algorithm.
}
	
\newcommand{\maincor}{  
	Given the constraints $ A\cdot V = 0 $ and $ l \leq V \leq u $,	where $ A \in M_{m \times n} (\mathbb{R})$  and $ l,V,u \in \rn{n} $,
	exactly one of these two options holds:
	\begin{enumerate}
		\item The \sat{} case:
		$ \exists V \in \rn{n} $ such that $ A \cdot V =
		0 $ and $ l \leq V \leq u $. 
		\item The \unsat{} case: $ \exists w \in \rn{m} $ such
		that for all  $ l \leq V \leq u $,  $ w^\intercal
                \cdot A \cdot V < 0$, whereas $ 0 \cdot w = 0 $. Thus,
                $w$ is a proof of the constraints' unsatisfiability.
              \end{enumerate}
 	Moreover, these vectors can be constructed during the run of the Simplex algorithm.
}

\newtheorem{theorem}{Theorem}[]
\newtheorem{lemma}[theorem]{Lemma}

\title{Neural Network Verification with Proof Production}

\author{
  \IEEEauthorblockN{
    Omri Isac\IEEEauthorrefmark{1},
    Clark Barrett\IEEEauthorrefmark{2},
    Min Zhang\IEEEauthorrefmark{3}
    and
    Guy Katz\IEEEauthorrefmark{1}
  }
  \IEEEauthorblockA{
    \IEEEauthorrefmark{1}The Hebrew University of Jerusalem, 
    \IEEEauthorrefmark{2}Stanford University,
    \IEEEauthorrefmark{3}East China Normal University
  }
}

\begin{document}

\maketitle
  
\begin{abstract}
  Deep neural networks (DNNs) are increasingly being employed in
  safety-critical systems, and there is an urgent need to guarantee
  their correctness. Consequently, the verification community has
  devised multiple techniques and tools for verifying DNNs. When DNN
  verifiers discover an input that triggers an error, that is easy to
  confirm; but when they report that no error exists, there is no way
  to ensure that the verification tool itself is not flawed.  As
  multiple errors have already been observed in DNN verification
  tools, this calls the applicability of DNN verification into
  question.  In this work, we present a novel mechanism for enhancing
  Simplex-based DNN verifiers with \emph{proof production}
  capabilities: the generation of an easy-to-check witness of
  unsatisfiability, which attests to the absence of errors. Our proof
  production is based on an efficient adaptation of the well-known
  Farkas' lemma, combined with mechanisms for handling
  piecewise-linear functions and numerical precision errors.  As a
  proof of concept, we implemented our technique on top of the Marabou
  DNN verifier. Our evaluation on a safety-critical system for
  airborne collision avoidance shows that proof production
  succeeds in almost all cases and requires only minimal overhead.
 \end{abstract}

\section{Introduction}
\label{sec:Introduction}
Machine learning techniques, and specifically deep neural networks
(DNNs), have been achieving groundbreaking results in solving
computationally difficult problems. Nowadays, DNNs are
state-of-the-art tools for performing many safety-critical tasks in
the domains of healthcare~\cite{EsRoRaKuDeChCuCoThDe19},
aviation~\cite{JuKoOw19} and autonomous
driving~\cite{BoDeDwFiFlGoJaMoMuZhZhZhZi16}. DNN training is performed
by adjusting the parameters of a DNN to mimic a highly
complex function over a large set of input-output examples (the
training set) in an automated way that is mostly opaque to humans.

The Achilles heel of DNNs typically lies in generalizing their
predictions from the finite training set to an infinite input
domain. First, DNNs tend to produce unexpected results on inputs that
are considerably different from those in the training set; and second,
the input to the DNN might be perturbed by sensorial imperfections, or
even by a malicious adversary, again resulting in unexpected and
erroneous results. These weaknesses have already been observed in many
modern DNNs~\cite{GoShSz14,SzZaSuBrErGoFe13}, and have even been demonstrated
in the real world~\cite{EyEvFeLiRaXiPrKoSo18} --- thus
hindering the adoption of DNNs in safety-critical settings.

In order to bridge this gap, in recent years, the formal methods
community has started devising techniques for DNN verification
(e.g.,~\cite{AkKeLoPi19, AvBlChHeKoPr19, BaShShMeSa19, FrChMaOsSe20,
	GeMiDrTsCHVe18, HeLo20, HuKwWaWu17, LyKoKoWoLiDa20, PuTa10,
	SaDuMo19, GaGePuVe19, TrBaXiJo20, WaPeWhYaJa18, ZhShGuGuLeNa20}, among many
others). Typically, DNN verification tools seek to prove that
outputs from a given set of inputs are contained within a safe
subspace of the output space, using various methods such as SMT
solving~\cite{AbKe17,BaTi18,DeBj11}, abstract
interpretation~\cite{GeMiDrTsCHVe18}, MILP solving~\cite{TjXiTe17},
and combinations thereof. Notably, many modern
approaches~\cite{KaHuIbJuLaLiShThWuZeDiKoBa19, LyKoKoWoLiDa20,
  MuMaSiPuVe22, TjXiTe17} involve a \emph{search} procedure, in which
the verification problem is regarded as a set of constraints. Then,
various input assignments to the DNN are considered in order to
discover a counter-example that satisfies these constraints, or to
prove that no such counter-example exists.

Verification tools are known to be as prone to errors as any other
program~\cite{JiRi21, ZhSuYaZhPuSu19}. Moreover, the search
procedures applied as part of DNN verification typically involve the repeated
manipulation of a large number of floating-point equations; this
can lead to rounding errors and numerical stability issues,
which in turn could potentially compromise the verifier's
soundness~\cite{BaLiJo21, JiRi21}. When the verifier discovers a
counter-example, this issue is perhaps less crucial, as the
counter-example can be checked by evaluating the DNN; but when the
verifier determines that no counter-example exists, this conclusion is
typically not accompanied by a witness of its correctness.

In this work, we present a novel proof-production mechanism for a
broad family of search-based DNN verification algorithms.  Whenever
the search procedure returns \unsat{} (indicating that no
counter-example exists), our mechanism produces a proof certificate
that can be readily checked using simple, external checkers. The proof
certificate is produced using a constructive version of Farkas'
lemma, which guarantees the existence of a witness to the
unsatisfiability of a set of linear equations --- combined with
additional constructs to support the non-linear components of a DNN,
i.e., its piecewise-linear activation functions.  We show how to
instrument the verification algorithm in order to keep track of its
search steps, and use that information to construct the proof with
only a small overhead.

For evaluation purposes, we implemented our proof-production technique
on top of the Marabou DNN
verifier~\cite{KaHuIbJuLaLiShThWuZeDiKoBa19}. We then evaluated our
technique on the ACAS Xu set of benchmarks for airborne collision
avoidance~\cite{JuLoBrOwKo16, KaBaDiJuKo21}. Our approach was
able to produce proof certificates for the safety of various ACAS Xu
properties with reasonable overhead ($5.7\%$ on
average). Checking the proof certificates produced by our approach was
usually considerably faster than dispatching the original
verification query.

The main contribution of our paper is in proposing a proof-production
mechanism for search-based DNN verifiers, which can substantially
increase their \emph{reliability} when determining
unsatisfiability. However, it also lays a foundation for a conflict-driven
clause learning (CDCL)~\cite{ZhMaMoMa01} verification scheme for DNNs,
which might significantly improve the performance of search-based
procedures (see discussion in Sec.~\ref{sec:Conclusion}).

The rest of this paper is organized as follows.  In
Sec.~\ref{sec:Background} we provide relevant background on DNNs,
formal verification, the Simplex algorithm, and on using Simplex for
search-based DNN verification. In
Sec.~\ref{sec:Overview},~\ref{sec:SimplexProof}
and~\ref{sec:TreeProof}, we describe the proof-production mechanism for
Simplex and its extension to DNN verification. Next, in
Sec.~\ref{sec:Complexity}, we briefly discuss complexity-theoretical
aspects of the proof production. Sec.~\ref{sec:Evaluation} details
our implementation of the technique and its evaluation. We then
discuss related work in Sec.~\ref{sec:Related} and conclude with
Sec.~\ref{sec:Conclusion}.

\section{Background}
\label{sec:Background}

\mysubsection{Deep Neural Networks.}  \textit{Deep neural networks}
(DNNs)~\cite{FoBeCu16} are directed graphs, whose nodes (neurons) are
organized into layers. Nodes in the first layer, called the
\textit{input layer}, are assigned values based on the input to the DNN; and then the
values of nodes in each of the subsequent layers are computed
as functions of the values assigned to neurons in the preceding layer.  More
specifically, each node value is computed by first applying an affine
transformation to the values from the preceding layer and then
applying a non-linear \textit{activation function} to the result. The
final (output) layer, which corresponds to the output of the network,
is computed without applying an activation function.

One of the most common activation functions is the
\textit{rectified linear unit} (\relu), which is defined as:
\begin{center}
$ f(b) = \relu(b) = \begin{cases}
  b & b>0 \\
  0 & \text{otherwise.} \end{cases} $
\end{center}
When $b>0$, we say that the \relu{} is in the \textit{active} phase;
otherwise, we say it is in the \textit{inactive} phase. For
simplicity, we restrict our attention here to \relu{}s, although our
approach could be applied to other piecewise-linear functions (such as
\emph{max pooling}, \emph{absolute value}, \emph{sign}, etc.). Non piecewise-linear functions,
such as as \emph{sigmoid} or \emph{tanh}, are left for future work.

Formally, a DNN $ \mathcal{N}:\rn{m}\rightarrow\rn{k} $, is a sequence
of $ n $ layers $ L_0,...,L_{n-1} $ where each layer $ L_i $ consists of
$ s_i \in \mathbb{N} $ nodes, denoted $ v^1_i,...,v^{s_i}_i $. The
assignment for the
$ j^{th} $ node in the $ 1 \leq i < n-1 $ layer is computed as
\[
v^j_i = \relu{} \left( \underset{l=1} { \overset {s_{i-1}} { \sum } }
  w_{i,j,l} \cdot v^l_{i-1} + p^j_i \right) \]
  and neurons in the output layer are computed as: 
\[ v^j_{n-1} = \underset{l=1} { \overset {s_{n-2}} { \sum } }
w_{n-1,j,l} \cdot v^l_{n-2} + p^j_{n-1} \]
where
$ w_{i,j,l}$ and $ p^j_i $ are (respectively) the predetermined weights and biases of $ \mathcal{N} $.
 We set $s_0 = m $ and treat $ v^1_0,...,v^m_0 $ as the input of
$ \mathcal{N} $.

\begin{figure}[h]

  \begin{center}

  \def\layersep{1.2cm}
  \begin{tikzpicture}[shorten >=1pt,->,draw=black!50, node distance=\layersep,font=\footnotesize]
    
    \node[input neuron] (I-1) at (0,-1) {$x_1$};
    \node[input neuron] (I-2) at (0,-3) {$x_2$};
    
    \node[hidden neuron] (H-1) at (\layersep, -2) {$v_1$};
    
    \node[hidden neuron] (H-2) at (2*\layersep,-2) {$v2$};
    
    \node[output neuron] (O-1) at (3*\layersep,-2) {$y$};
    
    \draw[nnedge] (I-1) --node[above,pos=0.4] {$1$} (H-1);
    \draw[nnedge] (I-2) --node[below,pos=0.4] {$-1$} (H-1);
    
    \draw[nnedge] (H-1) --node[below] {$-2$} (H-2);
    \draw[nnedge] (H-2) --node[below] {$1$} (O-1);
    
    \node[above=0.05cm of H-1] (b1) {$\relu{}$};
    \node[above=0.05cm of H-2] (b1) {$\relu{}$};
    
  \end{tikzpicture}
\end{center}
\caption{A toy DNN.}
\label{fig:toyDnn}
\end{figure}
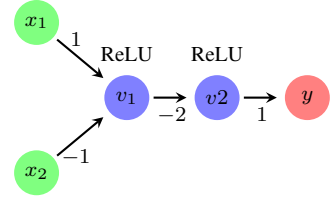
A simple DNN with four layers appears in Fig.~\ref{fig:toyDnn}. For
simplicity, the $ p_i^j $ parameters are all set to zero and are
ignored.  For input $\langle 1,2\rangle$, the node in the second layer
evaluates to
$ \relu(1 \cdot 1 \; + \; 2 \cdot (-1) ) = \relu(-1) = 0 $; the node
in the third layer evaluates to $ \relu(0 \cdot (-2)) = 0 $; and the
node in the fourth (output) layer evaluates to $ 0 \cdot 1 = 0$.

\mysubsection{DNN Verification and Proofs.} Given a DNN
$ \mathcal{N}:\rn{m}\rightarrow\rn{k} $ and a property
$ P:\rn{m+k}\rightarrow \lbrace \mathbb{T,F} \rbrace $, the \emph{DNN
  verification problem} is to decide whether there exist
$ x\in\rn{m}$ and $y\in\rn{k}$ such that
$ ( \mathcal{N}(x) = y )\land P(x,y) $ holds. If such $x$ and $y$ exist,
we say that the verification query $\langle \mathcal{N},P \rangle$ is
\textit{satisfiable} (\sat); and otherwise, we say that it is
\textit{unsatisfiable} (\unsat).  For example, given the toy DNN from
Fig.~\ref{fig:toyDnn}, we can define a property $P$:
$ P(x,y) \Leftrightarrow (x \in [2,3] \times [-1,1]) \wedge (y \in
[0.25,0.5]). $ Here, $P$ expresses the existence of an input
$x \in [2,3] \times [-1,1]$ that produces an output
$y \in [0.25,0.5] $.  Later on, we will prove that no such $x$ exists,
i.e., the verification query $\langle \mathcal{N},P \rangle$ is
\unsat{}.

Typically, $ P $ represents the negation of a desired property, and so
an input $x$ which satisfies the query is a counter-example --- whereas the
query's unsatisfiability indicates that the property holds. In this
work, we follow mainstream DNN verification
research~\cite{LyKoKoWoLiDa20, WaPeWhYaJa18} and focus on properties
$P$ that are a conjunction of linear lower- and upper-bound constraints on the neurons of $x$
and $y$. It has been shown that even for such simple properties, and
for DNNs that use only the \relu{} activation function, the
verification problem is NP-complete~\cite{KaBaDiJuKo21}.

A \textit{proof} is a mathematical object that certifies a
mathematical statement. In case a DNN verification query is \sat{},
the input $x$ for which $P$ holds constitutes a proof of the query's
satisfiability. Our goal here is to generate proofs also for the
\unsat{} case, which, to the best of our knowledge, is a feature that
no DNN verifier currently supports~\cite{BaLiJo21}.

\mysubsection{Verifying DNNs via Linear Programming.}  \textit{Linear
  Programming} (LP) \cite{Da63} is the problem of optimizing
 a linear function over a given convex polytope. An LP
instance over variables $V = \left[x_1,\ldots,x_n\right]^\intercal\in \rn{n}$ contains an objective function
$ c \cdot V $ to be maximized, subject to the constraints
$A \cdot V = b$ for some
$ A \in M_{m \times n}( \mathbb{R}), b\in \rn{m}$, and
$l \leq V \leq u$ for some
$l, u \in (\mathbb{R}\cup \lbrace \pm \infty \rbrace)^{n}
$. Throughout the paper, we use $l(x_i)$
and $u(x_i)$, to refer to the lower and upper bounds
(respectively) of $x_i$. LP solving can also be used to check the
\emph{satisfiability} of constraints of the form
$(A\cdot V = b) \wedge (l\leq V \leq u)$.

The \textit{Simplex} algorithm~\cite{Da63} is a widely
used technique for solving LP instances.
It begins by creating a \emph{tableau}, which is equivalent to the
original set of equations $AV=b$.  Next, Simplex selects a certain subset of the variables,
$\basic\subseteq\{x_1,\ldots,x_n\}$, to act as the \emph{basic
  variables}; and the tableau is considered as representing each basic
variable $x_i\in\basic$ as a linear combination of non-basic
variables, $x_i = \sumallnbvars $. 
We use $ A_{i,j} $ to denote the coefficient of a variable $ x_j $ in
the tableau row that corresponds to basic variable $ x_i $. Apart from
the tableau, Simplex also maintains a variable assignment that
satisfies the equations of $A$, but which may temporarily violate the
bound constraints $l\leq V \leq u$. The assignment for a variable
$ x_i $ is denoted $ \alpha(x_i) $.

After initialization, Simplex begins searching for an assignment that
simultaneously satisfies both the tableau and bound constraints. This
is done by manipulating the set $\basic$, each time swapping a basic
and a non-basic variable. This alters the equations of $A$ by adding
multiples of equations to other equations, and allows the algorithm to
explore new assignments.  The algorithm can terminate with a \sat{}
answer when a satisfying assignment is discovered or an \unsat{}
answer when:
\begin{inparaenum}[(i)]
\item a variable has contradicting bounds, i.e., $l(x_i)>u(x_i)$; or
\item one of the tableau equations $x_i=\sumallnbvars$ implies that
  $x_i$ can never satisfy its bounds.
\end{inparaenum}
The Simplex algorithm is sound, and is also complete if certain
heuristics are used for selecting the manipulations of
$\basic$~\cite{Da63}.  A detailed calculus for the
version of Simplex that we use appears in
Appendix~\ref{app:SimplexCalculus}.

LP solving is particularly useful in the context of DNN verification,
and is used by almost all modern tools (either natively~\cite{KaBaDiJuKo21}, or by
invoking external solvers such as GLPK~\cite{Ma08} or
Gurobi~\cite{gurobi}). More specifically, a DNN verification query can be
regarded as an LP instance with bounded variables that represents the property $P$ and the
affine transformations within $\mathcal{N}$, combined with a set of
piecewise-linear constraints that represent the activation
functions. We demonstrate this with an example, and then explain how
this formulation can be solved.

\begin{figure}[ht]
\centering
\scalebox{0.75} {
  \def\minWidth{4.0cm}
  \def\minHeight{4.4cm}
    \begin{tikzpicture}[ >=stealth,shorten >=1pt,shorten <=1pt]

      \tikzstyle{fancytitle} =[fill=blue!50!yellow, text=white, ellipse]

      \node[draw, rectangle, rounded corners, align=center, minimum width = \minWidth,
      minimum height = \minHeight, fill = navy!40]
      (LP)
      {
        $e^1 : \; b_1=x_1-x_2$ \\
        $e^2 : \; b_2=-2f_1$ \\
        $e^3 : \; y=f_2$ \\
        \\
        $
        \begin{bmatrix}
          2 \\ -1 \\ 0 \\ 0.25 \\ -0.5
        \end{bmatrix}
        \leq
        \begin{bmatrix}
          x_1 \\
          x_2 \\
          f_1 \\
          f_2, y \\
          b_1,b_2 
        \end{bmatrix}
        \leq
        \begin{bmatrix}
          3 \\
          1 \\
          0.5 \\
          0.5 \\
          0.5
        \end{bmatrix}
        $
      };

      \node[fancytitle] at ($(LP.north) + (-0cm,0cm)$) {Linear};

      \node[draw, rectangle, rounded corners, align=center, minimum width = \minWidth,
      minimum height = \minHeight, fill = red!40]
      (PLC) [right=1cm of LP]
      {
        \\
        $f_1=\relu(b_1)$ \\
        $f_2=\relu(b_2)$ \\
        \\
        \\
        \\
        \\
        \\
      } ;

      \node[fancytitle] at ($(PLC.north) + (-0cm,0cm)$) {Piecewise-Linear};

      \tikzstyle{background_rectangle}=[rounded corners, fill =
      navy!10]

      \begin{pgfonlayer}{background}

        \draw[background_rectangle]
        ($(LP.south west)
        + (-0.5cm, -0.2cm ) $)
        rectangle 
        ($(PLC.north east) + ( 0.5cm, 0.6cm) $);

        \coordinate (middle) at ($(LP.north)!.5!(PLC.north)$);
        \node[fancytitle] at ($(middle) + (0cm, 0.6cm)$) {Query $\varphi$};

      \end{pgfonlayer}
      
    \end{tikzpicture}
  }
  \caption{An example of a DNN verification query $\varphi$, comprised
    of an LP instance and piecewise-linear constraints.}
  \label{fig:query}
\end{figure}
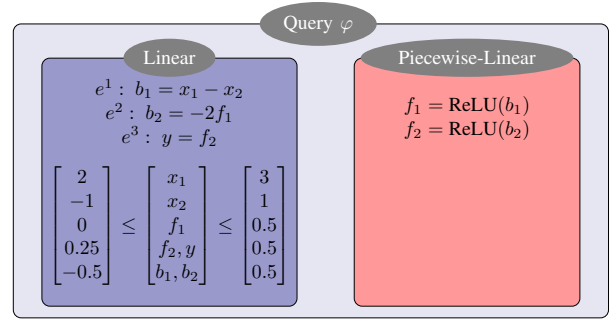

Recall the toy DNN from Fig.~\ref{fig:toyDnn}, and property $P$ that
is used for checking whether there exists an input $x$ in the range
$ [2,3] \times [-1,1]$ for which $\mathcal{N}$ produces an output $y$
in the range $ [0.25,0.5] $. We use $b_1, f_1$ to denote the input and
output to node $v_1$; $b_2,f_2$ for the input and output of $v_2$;
$x_1$ and $x_2$ to denote the network's inputs, and $y$ to denote the
network's output.  The linear constraints of the network yield the
linear equations $ b_1 = x_1 - x_2 $, $ b_2 = -2f_1 $, and $y = f_2$
(which we name $e^1,e^2$, and $e^3$, respectively).
The restrictions on the network's input and output are translated to
lower and upper bounds:
$ 2 \leq x_1 \leq 3, \: -1 \leq x_2 \leq 1, \: 0.25 \leq y \leq 0.5$.
The third equation implies that  $0.25 \leq f_2 \leq 0.5 $, which in turn implies that $b_2 \leq 0.5$. 
Assume we also restrict:
$ -0.5 \leq b_2, -0.5 \leq b_1 \leq 0.5, \: 0 \leq f_1 \leq 0.5, \:$.
Together, these constraints give rise to the linear program
that appears in Fig.~\ref{fig:query}.
The remaining \relu{} constraints, i.e. $ f_i = \relu(b_i) $
for $ i \in \lbrace 1,2 \rbrace $, exist alongside the LP instance.
Together, query $\varphi$ is equivalent to the DNN verification
problem that we are trying to solve.

Using this formulation, the verification problem can be solved using
Simplex, enhanced with a \emph{case-splitting} approach for handling
the \relu{} constraints~\cite{BaIoLaVyNoCr16,
  KaBaDiJuKo21}. Intuitively, we first invoke the LP solver on the LP
portion of the query; and if it returns \unsat{}, the whole query is
\unsat{}. Otherwise, if it finds a satisfying assignment, we check
whether this assignment also satisfies the \relu{} constraints.  If it
does, then the whole query is \sat{}. Otherwise, case splitting is
applied in order to split the query into two different sub-queries,
according to the two phases of the \relu{} function.\footnote{The
  approach is easily generalizable to any piecewise-linear constraint,
  by splitting the query according to the different linear pieces of the activation function.}  
Specifically, in one of the sub-queries, the LP query
is adjusted to enforce the \relu{} to be in the active phase: the
equation $f=b$ is added, along with the bound $b\geq 0$. In the other
sub-query, the inactive phase is enforced: $b\leq 0, 0\leq f\leq 0$.
This effectively reduces the \relu{} constraint into linear
constraints in each sub-query. This process is then repeated for each
of the two sub-queries.

Case-splitting turns the verification procedure into a \emph{search
  tree}~\cite{KaBaDiJuKo21}, with nodes corresponding to the splits
that were applied.  The tree is constructed iteratively, with Simplex
invoked on every node to try and derive \unsat{} or find a true
satisfying assignment. If Simplex is able to deduce that all leaves in
the search tree are \unsat{}, then so is the original
query. Otherwise, it will eventually find a satisfying assignment that
also satisfies the original query. This process is sound, and will
always terminate if appropriate splitting strategies are
used~\cite{Da63, KaBaDiJuKo21}.
Unfortunately, the size of the search tree can be exponential in the
number of \relu{} constraints; and so in order to keep the search tree
small, case splitting is applied as little as possible, according to
various heuristics that change from tool to tool~\cite{MuMaSiPuVe22,
  GaGePuVe19, WaPeWhYaJa18}. In order to reduce the number of splits even further, 
verification algorithms apply clever deduction techniques for
discovering tighter variable bounds, which may in turn rule out some
of the splits a-priori. We also discuss this kind of deduction, which
we refer to as dynamic bound tightening, in the following sections.

\section{Proof Production Overview}
\label{sec:Overview}

A Simplex-based verification process of a DNN is tree-shaped, and so
we propose to generate a \emph{proof tree} to match it. Within the
proof tree, internal nodes will correspond to case splits, whereas
each leaf node will contain a proof of unsatisfiability based on all
splits performed on the path between itself and the root.  Thus, a
proof tree constitutes a valid proof of unsatisfiability if each of
its leaves contains a proof that demonstrates that all splits so far
lead to a contradiction.  The proof tree might also include proofs for
\emph{lemmas}, which are valid statements for the node in which they
reside and its descendants (lemmas are needed for supporting bound
tightening, as we discuss later).

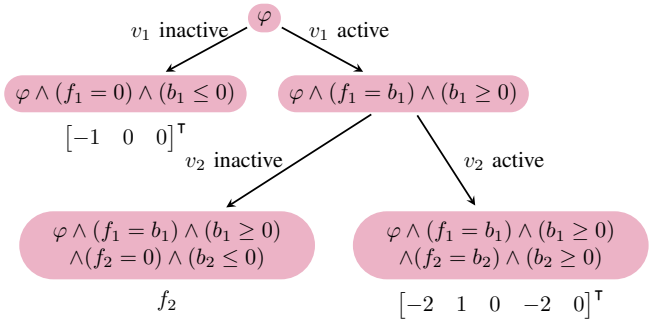
\begin{figure}[ht]
	\centering
	\scalebox{0.84} {
		\def\xSep{2.2cm}
		\def\ySep{1.2cm}
		\begin{tikzpicture}[ >=stealth,shorten >=1pt,shorten <=1pt]
			
			\node[proofNode] (root) at (0,0) []  {$\varphi$};
			\node[proofNode] (v1inactive) at (-\xSep, -\ySep)  [] {$\varphi\wedge (f_1=0)\wedge (b_1\leq 0)$};
			\node[proofNode] (v1active) at (\xSep, -\ySep)  []  {$\varphi\wedge (f_1=b_1)\wedge (b_1\geq 0)$};
			
			\node[proofNode] (v2inactive) at (-0.7*\xSep, -3*\ySep)  [] { 
				\begin{tabular}{c}
					$\varphi\wedge (f_1=b_1)\wedge (b_1\geq 0)$ \\ 
					$\wedge (f_2=0)\wedge (b_2\leq 0)$
				\end{tabular}
			};
			
			\node[proofNode] (v2active) at (1.7*\xSep, -3*\ySep)  [] {
				\begin{tabular}{c}
					$\varphi\wedge (f_1=b_1)\wedge (b_1\geq 0)$ \\ 
					$\wedge (f_2=b_2)\wedge (b_2\geq 0)$
				\end{tabular}
			};
			
			\draw[proofEdge] (root) --node[label={[xshift=-0.4cm,yshift=-0.1cm]$v_1$ inactive}] {} (v1inactive);
			\draw[proofEdge] (root) --node[label={[xshift=0.4cm,yshift=-0.1cm]$v_1$ active}] {} (v1active);
			
			\draw[proofEdge] (v1active) --node[label={[xshift=-1.0cm,yshift=-0.4cm]$v_2$ inactive}] {} (v2inactive);
			\draw[proofEdge] (v1active) --node[label={[xshift=0.9cm,yshift=-0.4cm]$v_2$ active}] {} (v2active);
			
			\node[below=0.05cm of v2inactive] {$f_2$};
			\node[below=0.05cm of v1inactive] {$\begin{bmatrix} -1 & 0 & 0 \end{bmatrix}^\intercal$};
			\node[below=0.05cm of v2active] {$\begin{bmatrix} -2 & 1 & 0 & -2
					& 0\end{bmatrix}^\intercal$};
			
		\end{tikzpicture}
	}
	\caption{A proof tree example.}
	\label{fig:example}
\end{figure}

As a simple, intuitive example, we depict in Fig.~\ref{fig:example} a
proof of unsatisfiability for the query $\varphi$ from
Fig.~\ref{fig:query}.  The root of the proof tree represents the
initial verification query, which is comprised of LP constraints and
\relu{} constraints. The fact that this node is not a leaf indicates
that the Simplex-based verifier was unable to conclude \unsat{} in
this state, and needed to perform a case split on the \relu{} node
$v_1$. The left child of the root corresponds to the case where
\relu{} $v_1$ is inactive: the LP is augmented with additional
constraints that represent the case split, i.e., $f_1=0$ and
$b_1\leq 0$. This new fact may now be used by the Simplex procedure,
which is indeed able to obtain an \unsat{} result. The node then
contains a proof of this unsatisfiability:
$\begin{bmatrix} -1 & 0 & 0 \end{bmatrix}^\intercal$. This vector
instructs the checker how to construct a linear combination of the
current tableau's rows, in a way that leads to a bound contradiction,
as we later explain in Sec.~\ref{sec:TreeProof}.

In the right child of the root, which represents $v_1$'s active phase,
the constraints $f_1 = b_1$ and $b_1 \geq 0$ are added by the
split. This node is not a leaf, because the verifier performed a
second case split, this time on $v_2$. The left child represents
$v_2$'s inactive phase, and has the corresponding constraints $f_2=0$ and
$b_2\leq 0$. This child is a leaf, and is marked with $f_2$,
indicating that $f_2$ is a variable whose bounds led to a
contradiction. Specifically, $f_2\geq 0.25$ from $\varphi$ and $f_2=0$
from the case split are contradictory.

The last node (the rightmost leaf) represents $v_2$'s active phase,
and has the constraints $f_2 = b_2$ and $b_2 \geq 0$. Here, the node
indicates that a contradiction can be reached from the current
tableau, using the vector
$\begin{bmatrix} -2 & 1 & 0 & -2 & 0\end{bmatrix}^\intercal$.  In
Sec.~\ref{sec:SimplexProof}, we explain how this process works.

Because each leaf of the proof tree contains a proof of
unsatisfiability, the tree itself proves that the original query is
\unsat{}. Note that many other proof trees may exist for the same
query.  In the following sections, we explain how to instrument a
Simplex-based verifier in order to extract such proof trees from the
solver execution.

\section{Simplex with Proofs}
\label{sec:SimplexProof}

\subsection{Producing proofs for LP}

We now describe our approach for creating proof trees, beginning with
leaf nodes. We start with the following lemma:

\begin{lemma}
  \label{thm:Lemma1}
  If Simplex returns \unsat{}, then there exists a variable with
  contradicting bounds; that is, there exists a variable 
  $ x_i \in V $ with lower and upper bounds $l(x_i)$ and $u(x_i)$, for
  which Simplex has discovered that  $ l(x_i) > u(x_i) $.
\end{lemma}

This lemma justifies our choice of using contradicting
bounds as proofs of unsatisfiability in the leaves of the proof tree.
The lemma follows directly from the derivation rules of
Simplex. Specifically, there are only two ways to reach \unsat{}: when
the input problem already contains inconsistent bounds
$ l(x_i) > u(x_i) $, or when Simplex finds a tableau row
$x_i = \sumallnbvars $ that gives rise to such inconsistent
bounds. The complete proof appears in Appendix~\ref{app:Lemma1Proof}.

We demonstrate this with an example, based on the query $\varphi$
from Fig.~\ref{fig:query}. Suppose that, as part of its Simplex-based
solution process, a DNN verifier performs two case splits,
fixing the two \relu{}s to their active states: $f_1=b_1\wedge b_1\geq
0$
and $f_2=b_2\wedge b_2\geq 0$. This gives rise to the following
(slightly simplified) system of equations:
\[
	 b_1 = x_1 - x_2 \quad b_2 = -2f_1 \quad y = f_2 \quad
	 f_1 = b_1 \quad f_2 = b_2
\]
Which corresponds to the tableau and variables
\[
  A =
  \begin{bmatrix}
    1 & -1 & -1 & 0  & 0  & 0  & 0  \\
    0 &  0 &  0 & -1 & -2 & 0 &  0 \\
    0 &  0 &  0 &  0 & 0  & 1  & -1 \\
    0 &  0 &  1 & 0  & -1 & 0  &  0 \\
    0 &  0 &  0 & 1  &  0 & -1 &  0 \\
    \end{bmatrix}\quad
  V = 
  \begin{bmatrix}
    x_1 \\ x_2 \\ b_1 \\ b_2 \\ f_1 \\ f_2 \\ y
  \end{bmatrix}^\intercal
\]
such that $AV=\overline{0}$, with the corresponding bound vectors:
\begin{align*}
	l &=
	\begin{bmatrix}
		2 & -1 & 0 \; \; \: & 0 \; \; \: & 0 \; \; \: & 0.25 & 0.25 
	\end{bmatrix}^\intercal \\
	u &=
	\begin{bmatrix}
		3 & 1 & 0.5 & 0.5 & 0.5 & 0.5 \; \: & 0.5
	\end{bmatrix}^\intercal
\end{align*}

Then, the Simplex solver iteratively alters the set of basic
variables, which corresponds to multiplying various equations by
scalars and summing them to obtain new equations. At some point, the
 equation $  b_2 = -2x_1 + 2x_2$ is obtained (by computing $\begin{bmatrix} -2 & 1 &
  0 & -2 & 0 \end{bmatrix}^\intercal \cdot A \cdot V$), with a current
assignment of
$
  \alpha(V)^{\intercal} = \begin{bmatrix}
    2 & 1 & 1 & -2 & 1 & -2 & -2
  \end{bmatrix}.
$

At this point, the Simplex solver halts with an \unsat{} notice. The
reason is that $b_2$ is currently assigned the value $-2$, which is
below its lower bound of $0$, and so its value needs to be
increased. However, the equation, combined with the fact that $x_1$ is pressed against its lower bound, while $x_2$ is pressed
against its upper bound, indicates that there is no slack remaining in
order to increase the value of $b_2$ (this corresponds to the \failureSlack{} rule
in the Simplex calculus in Appendix~\ref{app:SimplexCalculus}).  The key point is that the
same equation could be used in deducing a tighter bound for $b_2$:
\begin{align*}
  b_2 & \leq -2l(x_1) + 2u(x_2) 
       = -2\cdot 2 + 2\cdot 1 = -2
\end{align*}
and a contradiction could then be obtained based on the contradictory facts $0 = l(b_2) \leq b_2
\leq -2$. In other words, and as we formally prove in Appendix~\ref{app:Lemma1Proof},
any \unsat{} answer returned by Simplex can be regarded as a case of
conflicting lower and upper bounds.

Given Lemma~\ref{thm:Lemma1}, our goal is to instrument the Simplex
procedure so that whenever it returns \unsat{}, we are able to produce
a proof which indicates that $l(x_i)>u(x_i)$ for some variable $x_i$.
To this end, we introduce the following adaptation
of \textit{Farkas' Lemma}~\cite{Va96} to the
Simplex setting, which states that a linear-sized proof of this
fact exists.

\begin{lemma}
  \label{thm:maincor}
  \maincor
\end{lemma}
This Lemma is actually a corollary of Theorem~\ref{thm:mainthm}, which
we introduce later. For a complete proof, see
Appendix~\ref{app:MainCorProof}.

In our previous, \unsat{} example, one possible vector is
$w=\begin{bmatrix} -2 & 1 & 0 & -2 &
  0\end{bmatrix}^\intercal$. Indeed, $w\cdot A\cdot V=0$ gives us the equation
$ -2x_1 + 2x_2 - b_2 = 0$. Given the lower and upper bounds for the
participating variables, the
largest value that the left-hand side of the equation can obtain is:
\[
   -2l(x_1) + 2u(x_2) - l(b_2) 
  = -2 \cdot 2 + 2\cdot 1 - 0
   = -2 < 0
\]
Therefore, no variable assignment within the stated bounds can satisfy
the equation, indicating that the constraints are \unsat{}.

Given Lemma~\ref{thm:maincor}, all that remains is to instrument the
Simplex solver in order to produce the proof vector $w$ on the fly,
whenever a contradiction is detected. In case a trivial contradiction
$l(x_i) > u(x_i)$ is given as part of the input query for some
variable $x_i$, we simply return ``$x_i$'' as the proof (we later
discuss also how to handle this case in the presence of dynamic bound
tightenings). Otherwise, a non-trivial contradiction is detected as a
result of an equation $e\equiv x_i = \sumallnbvars$, which contradicts
one of the input bounds of $x_i$.  In this case, no assignment can
satisfy the equivalent equation $ \sumallnbvars - x_i = 0 $.  Since
the Simplex algorithm applies only linear operations to the input
tableau, $ e $ is given by a linear combination of the original
tableau rows.  Let $\coef(e)$ denote the Farkas vector of the equation
$e$, i.e., the column vector such that
$\coef(e)^\intercal \cdot A = e$, and which proves unsatisfiability in
this case. Our framework simply keeps track, for each row of the
tableau, of its coefficient vector; and if that row leads to a
contradiction, the vector is returned.

\subsection{Supporting dynamic bound tightening}
\label{sec:LPwithDynamicBounds}

So far, we have only considered Simplex executions that do not perform
any bound tightening steps; i.e., derive \unsat{} by finding a
contradiction to the original bounds. However, in practice, modern DNN
solvers perform a great deal of dynamic bound tightening, and so this
needs to be reflected in the proof.

We use the term \emph{ground bounds} to refer to variable bounds that
are part of the LP being solved, whether they were introduced by the
original input, or by successive case splits, as we will explain in
Sec.~\ref{sec:TreeProof}. This is opposed to \emph{dynamic bounds},
which are bounds introduced on the fly, via bound tightening. The
ground bounds, denoted $l,u\in \rn{n}$, are used in explaining dynamic
bounds, denoted $l',u'\in\rn{n}$, via Farkas vectors.

For simplicity, we consider here a simple and popular version of bound
tightening, called \emph{interval
  propagation}~\cite{Eh17,KaBaDiJuKo21}. Given an equation
$x_i = \sumallnbvars$ and current bounds $l'(x)$ and $u'(x)$ for each
of the variables (whether these are the ground bounds or dynamically
tightened bounds themselves), a new upper bound for $x_i$ can be
derived:
\begin{align}
  \label{eq:upperBoundUpdate}
  u'(x_i) := \underset{x_j\notin\basic, \: c_j > 0}{\sum }  c_j \cdot u'(x_j) 
  +  \underset{x_j\notin\basic , \: c_j < 0}{\sum }  c_j \cdot l'(x_j)
\end{align}
(provided that the new bound is tighter, i.e., smaller, than the
current upper bound for $x_i$).
A symmetrical version exists for discovering lower bounds.

A naive approach for handling bound tightening is to store, each time a
new bound is discovered, a separate proof that justifies it; for
example, a Farkas vector for deriving the equation that was used in
the bound tightening. However, a Simplex execution can include many
thousands of bound tightenings --- and so doing this would strain
resources. Even worse, many of the intermediate bound tightenings
might not even participate in deriving the final contradiction, and so
storing them would be a waste.

In order to circumvent this issue, we propose a scheme in which we
store, for each variable in the query, a single column vector that
justifies its current lower bound, and another for its current upper
bound. Whenever a tighter bound is dynamically discovered, the
corresponding vector is updated; and even if other, previously
discovered dynamic bounds were used in the derivation, the vector that we
store indicates how the same bound can be derived using the
ground bounds. Thus, the proof of the tightened
bounds remains compact, regardless of the number of derived bounds;
specifically, it requires only $ O( n \cdot m )$ space overall.
Formally, we have the following result:
\begin{theorem} 
\label{thm:mainthm}
\mainthm
\end{theorem}
When a Simplex procedure with bound
tightening reaches an \unsat{} answer, it has discovered a variable
$x_i$ with $ l'(x_i) > u'(x_i) $.  The theorem guarantees that in this case
we have two column vectors, $\farkasUpper{x_i}$ and
$\farkasLower{x_i}$, which \emph{explain} how $u'(x_i)$ and $l'(x_i)$
were discovered. We refer to these vectors as the \textit{Farkas vectors}
of the upper and lower bounds of $ x_i $, respectively.  Because
$u'(x_i) - l'(x_i)$ is negative, the column vector
$ w =\farkasUpper{x_i} - \farkasLower{x_i}$ creates a tableau row which
is always negative, making $ w \in\rn{m} $ a proof of unsatisfiability.
The formal, constructive proof of the theorem appears in
Appendix~\ref{app:MainThmProof}.

In order to maintain $\farkasUpper{x_i}$ and
$\farkasLower{x_i}$ during the execution of Simplex, 
whenever a tigher upper bound is tightened using
Eq.~\ref{eq:upperBoundUpdate}, 
we update the matching Farkas vector:
\[ \farkasUpper{x_i} := \sumposvars{ \farkasUpper{x_j} } + \sumnegvars{\farkasLower{x_j}} + \coef(e), \]
where $ e $ is the linear equation used for tightening, and
$ \coef(e) $ is the column vector such that
$\coef(e)^\intercal \cdot A = e$. The lower bound case is symmetrical.
To demonstrate the procedure, consider again the verification query
from Fig.~\ref{fig:query}. Assume the phases of $v_1, v_2$ have both been set to
active, and that consequently two new equations have been added:
$ e^4 : \; f_1 = b_1, \; e^5 : \; f_2 = b_2 $.  In this example, we have five linear equations,
so we initialize a zero vector of size five for each of the variable
bounds. Now, suppose Simplex tightens the lower bound of $b_1$ using
the first equation $e^1$:
\[
  l'(b_1) := l(x_1) - u(x_2) = 2 -1 = 1
\]
and thus
we update
\begin{align*}
 \farkasLower{b_1} &:= \farkasLower{x} - \farkasUpper{y} + \coef(e^1) \\
  &=
  \begin{bmatrix} 0 & 0 & 0 & 0 & 0 \end{bmatrix}^\intercal +
  \begin{bmatrix} 0 & 0 & 0 & 0 & 0\end{bmatrix}^\intercal \\
  & +\begin{bmatrix} 1 & 0 & 0 & 0 & 0\end{bmatrix}^\intercal \\
  &=
  \begin{bmatrix} 1 & 0 & 0 & 0 & 0\end{bmatrix}^\intercal
\end{align*}
since all $f_l$ and $f_u$ vectors have been initialized to
$\overline{0}$ and
$ \coef(e) = \begin{bmatrix} 1 & 0 & 0 & 0 & 0\end{bmatrix}^\intercal$
--- which indicates that $e^1$ is simply the first row of the
tableau.

We can now tighten bounds again, using the fourth row  $ f_1 = b_1$,
and get $ l'(f_1) := l'(b_1) = 1 $. We update $\farkasLower{f_1}$:
\begin{align*}
	\farkasLower{f_1} &:= \farkasLower{b_1} + \coef(e^4) \\
	&=
	 \begin{bmatrix} 1 & 0 & 0 & 0 & 0 \end{bmatrix}^\intercal +
	 \begin{bmatrix} 0 & 0 & 0 & 1 & 0\end{bmatrix}^\intercal \\
	&=\begin{bmatrix} 1 & 0 & 0 & 1 & 0\end{bmatrix}^\intercal
\end{align*}
To see that the Farkas vector can indeed explain the dynamically tightened
bound, observe that the combination $\begin{bmatrix} 1 & 0 & 0 & 1 & 0\end{bmatrix}^\intercal $ of tableau
rows gives the equation $ f_1 = x_1 - x_2 $. We can then tighten the lower bound
of $ f_1 $, using the \emph{ground} bounds:
$l'(f_1) := l(x_1) - u(x_2) = 2 - 1 = 1 $.  This bound matches the one
that we had discovered dynamically, though we derived it using 
ground bounds only.

\section{DNN Verification with Proofs}  
\label{sec:TreeProof}

\subsection{Producing a proof-tree}
\label{sec:proofTreesNoTightening}

We now discuss how to leverage the results of
Sec.~\ref{sec:SimplexProof} in order to produce the entire proof
tree for an \unsat{} DNN verification query. Recall that the main
challenge lies in accounting for the piecewise-linear constraints,
which affect the solving process by introducing case-splits.

Each case split performed by the solver introduces a branching in the
proof tree --- with a new child node for each of the linear phases of
the constraint being split on --- and introduces new equations and
bounds.  In the case of \relu{}, one child node represents the active
branch, through the equation $f=b$ and bound $b\geq 0$; and another
represents the inactive branch, with $b\leq 0$ and $0\leq f\leq 0$.
These new bounds become the \textit{ground bounds} for this node:
their Farkas vectors are reset to zero, and
all subsequent Farkas vectors refer to these new bounds (as opposed to the ground
bounds of the parent node). A new node inherits any
previously-discovered dynamic bounds, as well as the Farkas vectors
that explain them, from its parent; these vectors remain valid, as
ground bounds only become tighter as a result of splitting (see
Appendix~\ref{app:MaintainCorrectness}).

For example, let us return to the query from Fig.~\ref{fig:query} and
the proof tree from Fig.~\ref{fig:example}. Initially,
 the solver decides to split on $v_1$.  This adds two new
children to the proof tree.  In the first child, representing the
inactive case, we update the ground bounds
$ u(b_1) := 0, \: u(f_1) := 0 $, and reset the corresponding Farkas
vectors $\farkasUpper{b_1}$ and $\farkasUpper{f_1}$ to
$\overline{0}$. Now, Simplex can tighten the lower bound of $b_1$
using the first equation $e^1$:
\[
l'(b_1) := l(x_1) - u(x_2) = 2 -1 = 1
\]
resulting in the the updated
$\farkasLower{b_1}=\begin{bmatrix} 1 & 0 & 0\end{bmatrix}^\intercal$,
as shown in Sec.~\ref{sec:SimplexProof}, where we use vectors of size
three since in this search state we have three equations.  Observe
this bound contradicts the upper ground bound of $b_1$, represented by
the zero vector. We can then use the vector
\[
  \farkasUpper{b_1} - \farkasLower{b_1} =
  \overline{0}
  - \begin{bmatrix}1 & 0 & 0\end{bmatrix}^\intercal
  = \begin{bmatrix} -1 & 0 & 0 \end{bmatrix}^\intercal
\]
as a proof for contradiction.  Indeed, the matrix $A'$, which is
obtained using the first three rows and columns of $A$ as defined in
Sec.~\ref{sec:Overview}, corresponds to the tableau before adding any
new equations. Observe that
$\begin{bmatrix} -1 & 0 & 0 \end{bmatrix}^\intercal \cdot A' \cdot V =
0$ gives the equation $-x_1 + x_2 + b_1 = 0$. Given the current ground
bounds, the largest value of the left-hand side is:
\[
-l(x_1) + u(x_2) + u(b_1) = -2 + 1 + 0 = -1
\]
which is negative, meaning that no variable assignment within these
bounds can satisfy the equation. This indicates that the proof node
representing $v_1$'s inactive phase is \unsat.

In the second child, representing $v_1$'s active case, we update the
ground bound $l(b_1) : =0 $ and the Farkas vector
$\farkasLower{b_1} := \overline{0}$.  We also add the equation
$e^4 : \; f_1 = b_1$. Next, the solver performs another split on
$v_2$, adding two new children to the tree.  In the first one
(representing the inactive case) we update the ground bounds
$ u(b_2) := 0, \: u(f_2) := 0 $, and reset the corresponding Farkas
vectors $\farkasUpper{b_2}$ and $\farkasUpper{f_2}$ to $\overline{0}$.
In this node, we have a contradiction already in the ground bounds,
since $ u(f_2) := 0 $ but $ l(f_2) := 0.25 $. The contradiction in
this case is comprised of a symbol for $ f_2 $.

We are left with proving \unsat{} for the last child, representing the
case where both \relu{} nodes $v_1, v_2$ are active.  For this node of
the proof tree, we update the ground bound $ l(b_2) := 0 $ and Farkas
vector $\farkasLower{b_2} := \overline{0}$, and add the equation
$e^5: \; f_2 = b_2$. Recall that previously, we learned the tighter
bound $ l'(f_1) = 1 $. With the same procedure as described in
Sec.~\ref{sec:SimplexProof}, we can update
$\farkasLower{f_1} = \begin{bmatrix} 1 & 0 & 0 & 1 &
  0\end{bmatrix}^\intercal$.  Now, we can use $e^2 : \; b_2=-2f_1$ to
tighten $ u'(b_2) := -2l'(f_1) = -2 $, and consequently update the
Farkas vector:
\begin{align*}
	\farkasUpper{b_2} &= -2 \cdot  \farkasLower{f_1} + \coef(e^2) \\
	&= -2 \cdot 
	\begin{bmatrix} 1 & 0 & 0 & 1 & 0 \end{bmatrix}^\intercal +
	\begin{bmatrix} 0 & 1 & 0 & 0 & 0 \end{bmatrix}^\intercal \\
	&= 
	\begin{bmatrix} -2 & 1 & 0 & -2 & 0 \end{bmatrix}^\intercal
\end{align*}
The bound $u'(b_2) = -2$, explained by $\begin{bmatrix} -2 & 1 & 0 &
	-2 & 0\end{bmatrix}^\intercal$ contradicts the
ground bound $l(b_2) = 0$ explained by the zero vector. Therefore, we
get the vector
\[
\begin{bmatrix} -2 & 1 & 0 & -2 & 0\end{bmatrix}^\intercal -
\overline{0} = \begin{bmatrix} -2 & 1 & 0 & -2 & 0\end{bmatrix}^\intercal
\]
as the proof of contraction for this node. 

\subsection{Bound tightenings from piecewise-linear constraints}

Modern solvers often use sophisticated methods~\cite{Eh17,
  KaHuIbJuLaLiShThWuZeDiKoBa19, GaGePuVe19} to tighten variable bounds
using the piecewise-linear constraints.  For example, if
$ f = \relu(b) $, then in particular $b \leq f$, and so
$u(b) \leq u(f)$. Thus, if initially $u(b)=u(f)=7$ and it is later
discovered that $u'(f)=5$, we can deduce that also $u'(b)=5$.  We show
here how such tightening can be supported by our proof framework,
focusing on some \relu{} tightening rules as specified in
Appendix~\ref{app:ReLUProof}. Supporting additional rules
should be similar.

We distinguish between two kinds of \relu{} bound tightenings. The
first are tightenings that can be explained via a Farkas vector; these
are handled the same way as bounds discovered using interval
propagation.
The second, more complex tightenings are those that cannot be
explained using an equation (and thus a Farkas vector).  Instead, we
treat these bound tightenings as \emph{lemmas}, which are added to the
proof node along with their respective proofs; and the bounds that
they tighten are introduced as ground bounds, to be used in
constructing future Farkas vectors. The proof for a lemma consists of
Farkas vectors explaining any current bounds that were used in
deducing it; as well as an indication of the tightening rule that was
used. The list of allowed tightening rules must be agreed upon
beforehand and provided to the checker; in
Appendix~\ref{app:ReLUProof}, we present the tightening rules for
\relu{}s that we currently support.

For example, if $ f = \relu(b) $ and
$ u'(f) = 5 $ causes a bound tightening $ u'(b) = 5 $, then this new
bound $u'(b) = 5$ is stored as a lemma. Its proof consists of the
Farkas vector $\farkasUpper{f}$ which explains why $ u'(f) = 5 $, and
an indication of the deduction rule that was used (in this case,
$u'(b)\leq u'(f)$).

\section{Proof Checking and Numerical Stability}

 Checking the validity of a proof tree is
 straightforward. First, the checker must read the initial
query and confirm that it is consistent with the LP and
piecewise-linear constraints stored at the root of the tree. Next, the
checker begins a depth-first traversal of the proof tree. Whenever it
reaches a new inner node, it must confirm that that node's children
correspond to the linear phases of a piecewise-linear constraint
present in the query. Further, the checker must maintain a list of
current equations and lower and upper bounds, and whenever a new node
is visited --- update these lists (i.e., add equations and tighten
bounds as needed), to reflect the LP stored in that
node. Additionally, the checker must confirm the validity of lemmas
that appear in the node --- specifically, to confirm that they
adhere to one of the permitted derivation rules.  Finally, when a leaf
node is visited, the checker must confirm that the Farkas vector
stored therein does indeed lead to a contradiction when applied to the
current LP --- by ensuring that the linear combination of rows created
by the Farkas vector leads to a matrix row $\sum c_j \cdot x_j = 0$,
such that for any assignment of the variables, the left-hand side will have a negative value.

The process of checking a proof certificate is thus much simpler than
verifying a DNN using modern approaches, as it consists primarily of
traversing a tree and computing linear combinations of the tableau's
columns.  Furthermore, the proof checking process does not require
using division for its arithmetic computations, thus making the
checking program more stable arithmetically~\cite{JiRi21}.
Consequently, we propose to treat the checker as a trusted code-base,
as is commonly done~\cite{BaDeFo15, KaBaTiReHa16}.

\mysubsection{Complexity and Proof Size.}
\label{sec:Complexity}
Proving that a DNN verification query is \sat{} (by providing a
satisfying assignment) is significantly easier than discovering an
\unsat{} witness using our technique. Indeed, this is not surprising;
recall that the DNN verification problem is NP-complete, and that
yes-instances of NP problems have polynomial-size witnesses (i.e.,
polynomial-size proofs).  Discovering a way to similarly produce
polynomial proofs for no-instances of DNN verification is equivalent
to proving that  NP $=$ coNP, which is a major open
problem~\cite{ArBa09} and might, of course, be
impossible.

\mysubsection{Numerical Stability.}
Recall that enhancing DNN verifiers with
proof production is needed in part because they might produce incorrect
\unsat{} results due to numerical instability. When this happens, the
proof checking will fail when checking a proof leaf, and the user will receive warning. There
are, however, cases where the query is \unsat{}, but only the proof
produced by the verifier is flawed. 
To recover from these cases and correct the proof, we propose to use
an external SMT solver to re-solve the query stored in the leaf in question.

SMT solvers typically use sound arithmetic (as opposed to DNN
verifiers), and so their conclusions are generally more reliable. Further,
if a proof-producing SMT solver is used, the proof that it produces
could be plugged into the larger proof tree, instead of the incorrect
proof previously discovered. Although using SMT solvers to directly
verify DNNs has been shown to be highly
ineffective~\cite{KaBaDiJuKo21, PuTa12}, in our evaluation we observed
that leaves typically represented problems that were significantly
simpler than the original query, and could be solved efficiently by
the SMT solver.

\section{Implementation and Evaluation}
\label{sec:Evaluation}

\mysubsection{Implementation.}  For evaluation purposes, we
instrumented the Marabou DNN
verifier~\cite{KaHuIbJuLaLiShThWuZeDiKoBa19, WuOzZeIrJuGoFoKaPaBa20}
with proof production capabilities. Marabou is a state-of-the-art DNN
verifier, which uses a native Simplex solver, and combines it with
other modern techniques --- such as abstraction and abstract
interpretation~\cite{ElCoKa22, ElGoKa20, OsBaKa22, GaGePuVe19,
	WaPeWhYaJa18, ZeWuBaKa22}, advanced splitting heuristics~\cite{WuZeKaBa22}, DNN
optimization~\cite{StWuZeJuKaBaKo21}, and support for varied
activation functions~\cite{AmWuBaKa21}.  Additionally, Marabou has
been applied to a variety of verification-based tasks, such as
verifying recurrent networks~\cite{JaBaKa20} and DRL-based
systems~\cite{AmCoYeMaHaFaKa22, AmScKa21, ElKaKaSc21, KaBaKaSc19},
network repair~\cite{GoAdKeKa20, ReKa22}, network
simplification~\cite{GoFeMaBaKa20, LaKA21}, and ensemble
selection~\cite{AmKaSc22}.

As part of our enhancements to Marabou's Simplex core, we added a
mechanism that stores, for each variable, the current Farkas
vectors that explain its bounds. These vectors are updated with each
Simplex iteration in which the tableau is altered. Additionally, we
instrumented some of Marabou's Simplex bound propagation mechanisms
--- specifically, those that perform interval-based bound tightening
on individual rows~\cite{Eh17}, to record for each tighter bound the
Farkas vector that justifies it.  Thus, whenever the Simplex core
declares \unsat{} as a result of conflicting bounds, the proof
infrastructure is able to collect all relevant components for creating
the certificate for that particular leaf in the proof tree. Due to
time restrictions, we were not able to instrument all of Marabou's
many bound propagation components; this is ongoing work, and our
experiments described below were run with yet-unsupported components turned
off. The only exception is Marabou's preprocessing component, which is
not supported, but is run before proof production starts.

In order to keep track of Marabou's tree-like search, we instrumented
Marabou's \emph{SmtCore} class, which is in charge of case splitting
and backtracking~\cite{KaHuIbJuLaLiShThWuZeDiKoBa19}. Whenever a
case-split was performed, the corresponding equations and bounds were
added to the proof tree as ground truths; and whenever a previous
split was popped, our data structures would backtrack as well,
returning to the previous ground bounds.

In addition to the instrumentation of Marabou, we also wrote a simple
proof checker that receives a query and a proof artifact --- and then
checks, based on this artifact, that the query is indeed
\unsat{}. That checker also interfaces with the cvc5 SMT
solver~\cite{BaBaBrKrLaMaMoMoNiNoOzPrReShTiZo22} for attempting recovery from numerical instability
errors. 

\mysubsection{Evaluation.}  We used our proof-producing version of
Marabou to solve queries on the ACAS-Xu family of benchmarks for
airborne collision avoidance~\cite{JuKoOw19}. We argue that the
safety-critical nature of this system makes it a prime candidate for
proof production. Our set of benchmarks was thus comprised of 45
networks and 4 properties to test on each, producing a total of 180
verification queries. Marabou returned an \unsat{} result on 113 of
these queries, and so we focus on them. In the future, we intend to 
evaluate our proof-production mechanism on other benchmarks as well.

We set out to evaluate our proof production mechanism along 3 axes:
\begin{inparaenum}[(i)]
	\item \emph{correctness}: how often was the checker able to verify the
	proof artifact, and how often did Marabou (probably due to numerical
	instability issues) produce incorrect proofs?;
	\item \emph{overhead}: by how much did Marabou's runtime increase due
	to the added overhead of proof production?; and 
	\item \emph{checking time}: how long did it take to check the produced
	proofs?
\end{inparaenum}
Below we address each of these questions.

\begin{figure}[h]
	\begin{center}
		\scalebox{0.8}{
			\includegraphics[width=\linewidth]{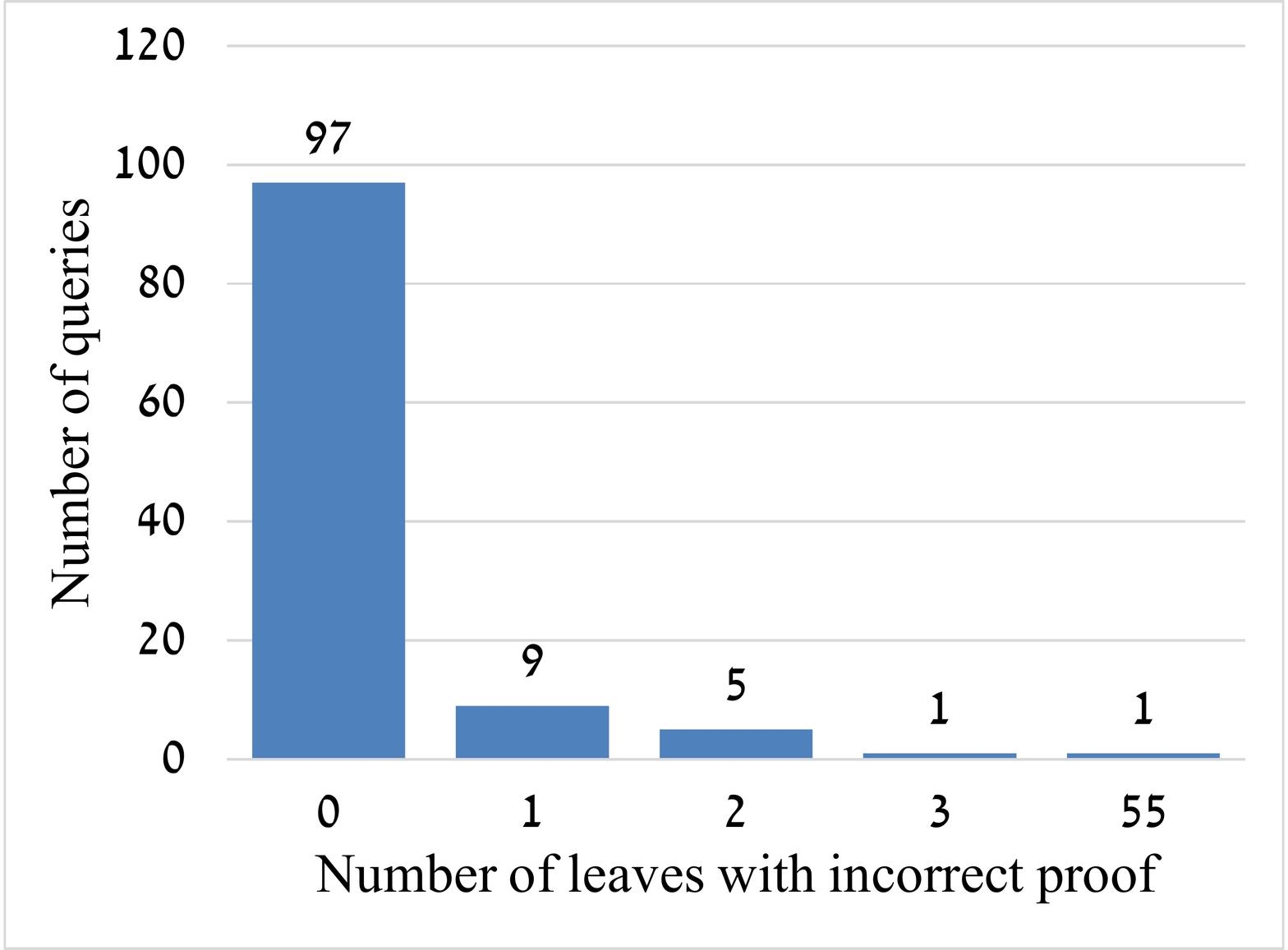}
		}
	\end{center}
	\caption{Number of queries per number of leaves with incorrect
		proofs.}
	\label{fig:delegation}
\end{figure}

\emph{Correctness.}  Over 1.46 million proof-tree leaves were created
and checked as part of our experiments. Of these,
proof checking failed for only 
77 leaves, meaning that the Farkas vector written in the proof-tree
leaf did not allow the proof checker to deduce a contradiction.
Out of the 113 queries checked,
97 had all their proof-tree leaves checked successfully. As for the
rest, typically only a tiny number of leaves would fail per query, but
we did identify a single query where a significant number of proofs
failed to check (see Fig.~\ref{fig:delegation}). We speculate that
this query had some intrinsic numerical issues encoded into it
(e.g., equations with very small coefficients~\cite{Ch83}).

Next, when we encoded each of the 77 leaves as a query to the cvc5 SMT
solver~\cite{BaBaBrKrLaMaMoMoNiNoOzPrReShTiZo22}, it was able to show that all queries
were indeed \unsat{}, in under 20 seconds per query. From this we
learn that although some of the proof certificates produced by Marabou
were incorrect, the ultimate \unsat{} result was correct. Further, it
is interesting to note how quickly each of the queries could be
solved. This gives rise to an interesting verification strategy: use
modern DNN verifiers to do the ``heavy-lifting'', and then use more
precise SMT solvers specifically on small components of the query that
proved difficult to solve accurately.

\begin{figure*}[t]
	\centering
	\begin{multicols}{2}
		\scalebox{0.85}{\includegraphics[width=\linewidth]{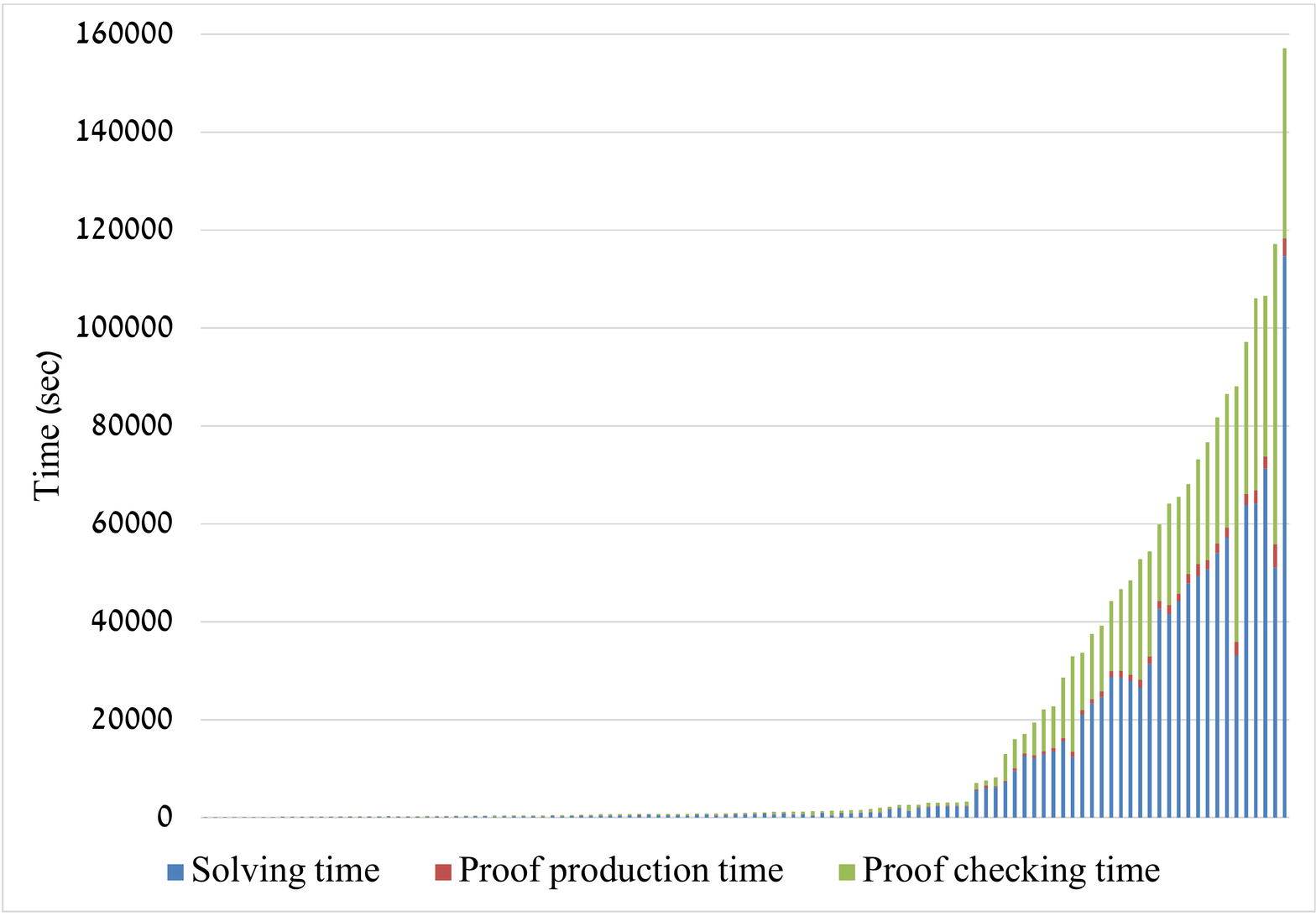}} 	 \par
		
		\scalebox{0.85}{\includegraphics[width=\linewidth]{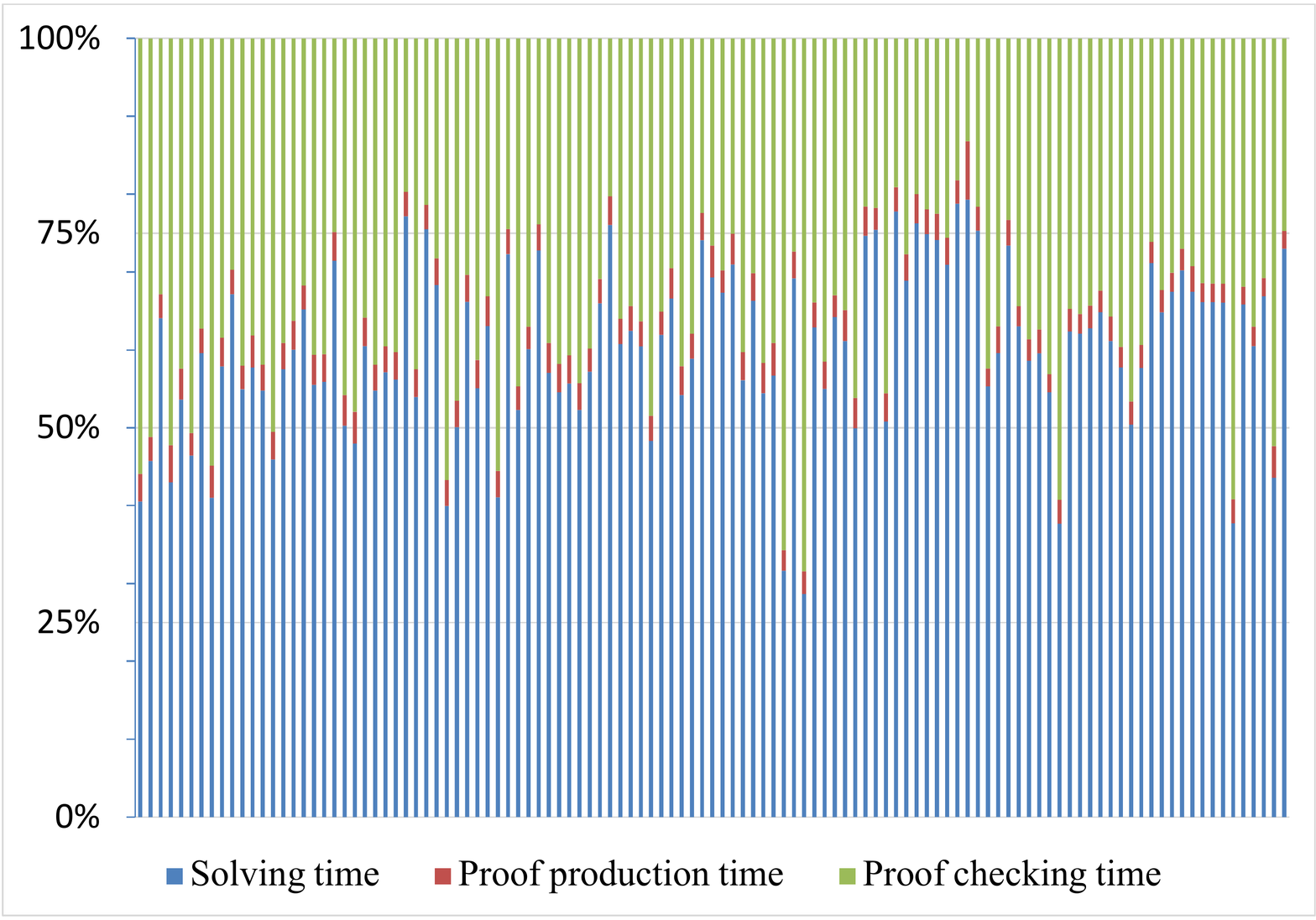}}
	\end{multicols}
	\caption{Proof production and checking time comparison --- absolute (left) and relative (right)}
	\label{fig:timecomparison}
\end{figure*}

\emph{Overhead and Checking Time.}  In Fig.~\ref{fig:timecomparison}, we compare the running time of vanilla Marabou,
the overhead incurred by our proof-production extension to Marabou,
and the checking time of the resulting proof certificates.  We can see
that the overhead of proof production time is relatively small for all
queries (an average overhead of $5.7\%$), while the certification
time is non-negligible, but shorter than the time it takes to solve
the queries by a factor of $66.5\%$ on average.

\section{Related Work}
\label{sec:Related}

The importance of proof production in verifiers has been repeatedly
recognized, for example by the SAT, SMT, and model-checking communities
(e.g.,~\cite{BaDeFo15, CoMeZa15, GrRoTo21}). Although the risks posed
by numerical imprecision within DNN verifiers have been raised
repeatedly~\cite{BaLiJo21, JiRi21, KaBaDiJuKo21, KaBaDiJuKo17}, we are unaware of
any existing proof-producing DNN verifiers.

Proof production for various Simplex variants has been studied
previously~\cite{Ne98}. In~\cite{DuDe06}, Dutertre and de Moura study
a Simplex variant similar to ours, but without explicit
support for dynamic bound tightening. Techniques for producing Farkas
vectors have also been studied~\cite{AvKa04}, but again without
support for dynamic bound tightening, which is crucial in DNN
verification. Other uses of Farkas vectors, specifically in the
context of interpolants, have also been explored~\cite{AsBlHyFeSh20,
  BlHyAnKoSh19}.

Other frameworks for proof production for machine learning have also
been proposed~\cite{AnZhWuGr21, GoRoShYe21}; but these frameworks are
interactive, unlike the automated mechanism we present here.

\section{Conclusion and Future work}
\label{sec:Conclusion}

We presented a novel framework for producing proofs of
unsatisfiability for Simplex-based DNN verifiers.  Our framework
constructs a proof tree that contains lemma proofs in internal nodes
and unsatisfiability proofs in each leaf. The certificates of
unsatisfiability that we provide can increase the reliability of DNN
verification, particularly when floating-point arithmetic (which is
susceptible to numerical instability) is used.

We plan to continue this work along two orthogonal paths:
\begin{inparaenum}[(i)]
\item
  extend our mechanism to support
  additional steps performed in modern verifiers, such as
  preprocessing and additional abstract interpretation
  steps~\cite{LyKoKoWoLiDa20, GaGePuVe19}; and
\item use our infrastructure to allow learning succinct \emph{conflict
    clauses}~\cite{ZhMaMoMa01}.  During search, the Farkas vectors produced by our
approach could be used to generate conflict clauses on-the-fly.
Intuitively, conflict clauses guide the verification algorithm
to avoid any future search for a satisfying assignment within
subspaces of the search space already proven to be \unsat. 
Such clauses are a key component in
modern SAT and SMT solvers, and are the main component of CDCL algorithms~\cite{ZhMaMoMa01} --- and could significantly curtail the
search space traversed by DNN verifiers and improve their
scalability.
\end{inparaenum}

\mysubsection{Acknowledgments.}  This work was supported by the Israel
Science Foundation (grant number 683/18), the ISF-NSFC joint research
program (grant numbers 3420/21 and 62161146001), the Binational
Science Foundation (grant numbers 2017662 and 2020250), and the
National Science Foundation (grant number 1814369).

\bibliographystyle{abbrv}
\bibliography{proof_production}

\newpage

{\noindent\huge{Appendix}}

\renewcommand\thesection{\Alph{section}}
\renewcommand\thesubsection{\thesection.\arabic{subsection}}
\setcounter{section}{0}
\renewcommand{\theHsection}{Supplement.\thesection}

\section{The Simplex Calculus}
\label{app:SimplexCalculus}

There exist numerous flavors and implementations of the Simplex
algorithm for solving LP instances~\cite{Da63}. To make
our proof-production scheme compatible with many of them, we used it
to extend the following, abstract version of Simplex, where the
algorithm is given as a set of derivation rules.  This formulation,
which appears below, is borrowed from Katz et
al.~\cite{KaBaDiJuKo21} (see therein for additional details).

Let $\allvars$ denote a set of variables:
$\allvars = \{x_1,\ldots,x_n\}$.  We define a simplex configuration to
be either one of the distinguished symbols $\{\sat{},\unsat{}\}$, or a
tuple $\langle \basic, A, \lb, \ub, \assignment\rangle$. The tuple is
comprised of the following components:
\begin{itemize}
\item $\basic\subseteq \allvars$ is the set of basic variables.
\item $A$, also called the
  \emph{tableau}, contains for each $x_i\in\basic$ an equation
  $x_i = \sum_{x_j\notin\basic} c_j x_j$.
\item
  $\lb$ and $\ub$ are mappings, which
assign to each variable $x\in\allvars$ a lower and an upper bound,
respectively.
\item $\assignment$, called the \emph{assignment}, is a mapping from
  each 
  variable $x\in\allvars$ to a real value assigned to it
\end{itemize}
When beginning a Simplex execution, the initial configuration
(including the initial tableau $A_0$) is created to match the input LP
instance. Specifically, each equation
$\sum_{x_i\in \allvars}c_ix_i = d$ that appears in the input is
assigned a fresh basic variable $b$, and the equation
$b=\sum_{x_i\in \allvars}c_ix_i$ is added to the
tableau. Additionally, $d$ is set as both the lower and upper bound
for $b$.  The initial assignment is $0$ for each of the variables,
thus ensuring that all tableau equations hold (though this assignment
may violate variable bounds).

It is often convenient to regard the tableau $A$ as a matrix, where
each row expresses a basic variables (a variable in $\basic{}$) as a
linear combination of non-basic variables (those in
$\allvars{}\setminus\basic{}$).

\begin{figure*}[t]
	\begin{centering}
		
		\pivot{1}
		\drule{
			x_i\in\basic,
			\ \ 
			\assignment (x_i) < \lb(x_i), 
			\ \ 
			x_j\in\slackPlus(x_i)
		}
		{
			A := \pivotOperation{}(A,i,j), 
			\ \ 
			\basic := \basic\cup \{x_j\} \setminus \{x_i\}
		}
		\medskip
		
		\pivot{2}
		\drule{
			x_i\in\basic,
			\ \ 
			\assignment (x_i) > \ub(x_i),
			\ \
			x_j\in\slackMinus(x_i)
		}
		{
			A:=\pivotOperation{}(A,i,j), 
			\ \ 
			\basic := \basic\cup \{x_j\} \setminus \{x_i\}
		}
		\medskip 
		
		\update
		\drule{
			x_j\notin\basic, 
			\ \ 
			\assignment(x_j) < \lb(x_j) \vee \assignment(x_j) > \ub(x_j), 
			\ \ 
			\lb(x_j) \leq \assignment(x_j) + \delta \leq \ub(x_j)
		}
		{
			\assignment := \updateOperation(\assignment, x_j, \delta)
		}
		\medskip
		
		\failureSlack
		\drule{
			x_i\in\basic,
			\ \ 
			( \assignment (x_i) < \lb(x_i)
			\ \wedge \
			\slackPlus(x_i)=\emptyset )
			\vee
			( \assignment (x_i) > \ub(x_i)
			\ \wedge \
			\slackMinus(x_i)=\emptyset )
		}
		{
			\unsat{}
		}
		\medskip
		
		\failureBounds
		\drule{
			\lb(x_i) > \ub(x_i)
		}
		{
			\unsat{}
		}
		\medskip
		
		\success
		\drule{
			\forall x_i\in\allvars. \ 
			\lb(x_i) \leq \assignment(x_i) \leq \ub(x_i)
		}
		{
			\sat{}
		}
		\caption{Derivation rules for the abstract simplex algorithm.}
		\label{fig:abstractSimplex}
	\end{centering}
\end{figure*}

Then, the rows of
$A$ correspond to the basic variables, and its columns to
the non-basic variables. For $x_i\in\basic{}$ and
$x_j\notin\basic{}$ we use $A_{i,j}$ to denote the coefficient
$c_j$ of $x_j$ in the equation $x_i=\sum_{x_j\notin\basic} c_j x_j$.
The tableau is changed via pivoting: the switching of one basic variable
$x_i$ (the \emph{leaving} variable) with one non-basic variable $x_j$
(the \emph{entering} variable), for which $A_{i,j}\neq 0$. When a
\pivotOperation{}($A,i,j$) operation is performed, it returns a new tableau, where the equation $x_i=\sum_{x_k\notin\basic}
c_k x_k$ has been replaced by the equation
$
x_j = \frac{x_i}{c_j} - \sum_{x_k\notin\basic, k\neq j}\frac{c_k}{c_j}x_k
$, and in which any occurrences $x_j$ in each of the other equations has been
replaced by the right-hand side of the new equation (and the resulting expressions
are normalized to fit the tableau form).
The variable assignment, $\assignment{}$, can be changed via
\emph{update} operations (which are
applied to non-basic variables):
for $x_j\notin\basic{}$, an \updateOperation{}($\assignment, x_j,\delta$) operation
returns
an updated
assignment $\assignment'$, which is identical to $\assignment$ except that
$\assignment'(x_j)=\assignment(x_j)+\delta$; and for
every $x_i\in \basic$, we have that
$
\assignment'(x_i)=\assignment(x_i)+\delta\cdot A_{i,j}.
$
To simplify later presentation we also denote:
\begin{align*}
\slack^+&(x_i)  =\\
& \{
x_j\notin\basic{}\ |\ 
(A_{i,j}>0
\wedge
\assignment(x_j) < \ub(x_j))
\vee \\ 
& (A_{i,j}<0
\wedge
\assignment(x_j) > \lb(x_j))
\}
\\
\slack^-&(x_i) = \\
& \{
x_j\notin\basic{}\ |\ 
(A_{i,j}<0
\wedge
\assignment(x_j) < \ub(x_j))
\vee \\
& (A_{i,j}>0
\wedge
\assignment(x_j) > \lb(x_j))
\}
\end{align*}

Fig.~\ref{fig:abstractSimplex} depicts the rules of the Simplex
calculus, in \emph{guarded assignment form}.  We say that a rule
applies to a configuration $S$ if all of its premises hold in $S$.  A
rule's conclusion describes how each component of $S$ is (possibly)
changed.  We say that $S$ derives $S'$ if $S'$ is the result of
applying a rule to $S$.  A \emph{derivation} is a sequence of
configurations $S_i$, where each $S_i$ derives $S_{i+1}$.

The Simplex calculus is known to be \emph{sound}~\cite{Va96} (i.e., if
a derivation ends in $\sat{}$ or $\unsat{}$, then the original problem
is satisfiable or unsatisfiable, respectively) and \emph{complete}
(there always exists a derivation ending in either $\sat{}$ or
$\unsat{}$ from any starting configuration).  Termination can be
guaranteed if certain strategies are used in applying the transition
rules --- in particular in picking the leaving and entering variables
when multiple options exist~\cite{Va96}.

\section{Proof of lemma~\ref{thm:Lemma1} }
\label{app:Lemma1Proof}

Recall the following lemma, which was introduced in
Section~\ref{sec:SimplexProof}: \setcounter{theorem}{0}
\begin{lemma}
  If Simplex returns \unsat{}, then there exists a variable with
  contradicting bounds; that is, there exists a variable 
  $ x_i \in V $ with lower and upper bounds $l(x_i)$ and $u(x_i)$, for
  which Simplex has discovered that  $ l(x_i) > u(x_i) $.
\end{lemma}
We now formally prove this lemma.

Suppose a Simplex run ended in the distinguished state \unsat{}.
There are precisely two deduction rules that result in a transition to
this state.
The first, \failureBounds{}, can only be triggered if there is an explicit
bound contradiction for one of the variable, thus proving the lemma. 
The second,  \failureSlack{},
can only be triggered if there exists a basic variable $ x_i $, whose
tableau row is 
$ x_i = \sumallnbvars  $,
with either
\[ \alpha(x_i) < l(x_i) \land slack^+(x_i) = \emptyset \]
or
\[ \alpha(x_i) > u(x_i) \land slack^-(x_i) = \emptyset \]

If the first condition holds, then variable $ x_i $'s assignment is
below its lower bound, and for all $ j \neq i $, if $ A_{i,j} < 0 $
(equivalently, $ c_j < 0 $), then $ x_j $ is pressed against its lower
bound; and if $ A_{i,j} > 0 $ (equivalently, $ c_j > 0 $), then
$ x_j $ is pressed against its upper bound.  Therefore, the right-hand
side of the equation $ x_i = \sumallnbvars $ is pressed against its
upper bound.  Since the upper bound of the left-hand side, $ x_i $,
cannot be greater than the upper bound of the right-hand side, we
conclude that:
\begin{align*}
  u( x_i ) &\leq
             \sum_{j\notin\basic, c_j>0}c_j\cdot u(x_j) +
             \sum_{j\notin\basic, c_j<0}c_j\cdot l(x_j) \\
           &=
             \sum_{j\notin\basic, c_j>0}c_j\cdot \alpha(x_j) +
             \sum_{j\notin\basic, c_j<0}c_j\cdot \alpha(x_j)  \\
           &= \alpha(x_i) < l(x_i)
\end{align*}
where the second transition is correct because the variables are pressed against their bounds, as we described.

The proof for the second case is symmetrical. In either case, we get
that if the \failureBounds{} rule is applicable, then Simplex has
discovered a variable with contradicting lower and upper bounds, as
needed.

\section{Proof of theorem~\ref{thm:mainthm} }
\label{app:MainThmProof}

Before proving the theorem, we begin by describing the procedure for
deducing a bound using a Farkas vector $ w $.
We assume without loss of generality, that the Tableau is of the form $AV = 0$, for $V$ the variables vector. If this is initially not so, we add a fresh variable
$x_j$ to every row $j$ of $A$, and set $l(x_j)=u(x_j)=-b_j$.

Given a tableau $ A \in M_{m \times n} $, bounds $ u,l $ and a column
vector $ w \in \rn{m} $, the row vector $ w^{\intercal} \cdot A \cdot V $
corresponds to an equation of the form
$ 0 = \underset{j = 0}{\overset{ n }{\sum} } c_j \cdot x_j $, or
equivalently $ x_i = \sumallvars + ( c_i + 1 ) \cdot x_i $.   We
seek to use the right-hand of the equation to deduce tighter bounds
for its left-hand side.

Suppose we are able to learn a tighter upper bound for $x_i$ (the
lower bound case is symmetrical).
In this case, the new upper bound learned for $x_i$ is the upper bound of
the expression
\[
  \sumallvars + ( c_i + 1 ) \cdot x_i,
\]
which is
\[
  \sumposvars{ u(x_j) } + \sumnegvars{ l(x_j) } + ( c_i + 1 ) \cdot \gamma(x_i)
\]
Where $\gamma(x_i) = u(x_i)$ if $c_i + 1 >0$, or $\gamma(x_i) = l(x_i)$ otherwise.

We now use the aforementioned observation in proving
Theorem~\ref{thm:mainthm}. Let us recall the theorem itself:

\setcounter{theorem}{2}
\begin{theorem} 
\mainthm
\end{theorem}

The proof that we provide next is constructive. Observe variable
$x_i$, for which a dynamic bound is currently being discovered. As we
observed before, this
tightening is being performed using an equation of the form
$ x_i = \sumallvars $,
which is equivalent to $ 0 = \sumallvars - x_i $.
We define the vector $ e = \begin{bmatrix} e_1, \ldots,
  e_n\end{bmatrix}$ as:
  \[
    e_k =
    \begin{cases}
      -1 & \text{if $k=i$} \\
      c_j & \text{otherwise}
      \end{cases}
    \]
    This vector represents the coefficients of the equation that is
    used for bound tightening using the product $e \cdot V$.

Suppose that the bound being tightened is $x_i$'s upper bound. In that
case, the new bound is given by:
\[
  u'(x_i) := \sumposvars{u'(x_j)} + \sumnegvars{l'(x_j)}
\]
i.e., the new upper bound of $x_i$ is set to match the upper bound of
the equation on the right-hand side.

Upon initialization of the Simplex algorithm, for each variable $ x_i $
we assign two vectors, $ \farkasUpper{x_i}, \farkasLower{x_i} =
\overline{0} $, to represent that no bound tightening has occurred for
$x_i$.  Whenever one of its  bounds is updated, the corresponding
vector is updated. In the case of an upper bound, we set

\[
  \farkasUpper{x_i} := \sumposvars{ \farkasUpper{x_j} } + \sumnegvars{
    \farkasLower{x_j} } + \coef(e)
\]
where $ \coef(e) $ is a column vector such that $ \coef(e)^\intercal \cdot A = e $.

The case of $x_i$'s lower bound is symmetrical. The tightening is
performed according to the equation
\[
  l'(x_i) := \sumposvars{l'(x_j)} + \sumnegvars{u'(x_j)}
\]
and the update rule is
\[
  \farkasLower{x_i} := \sumposvars{ \farkasLower{x_j} } + \sumnegvars{
    \farkasUpper{x_j} } + \coef(e)
\]

In order to prove the theorem, we need to prove the following
invariant: regardless of the bounds dynamically tightened so far, the
two vectors $ \farkasUpper{x_i}, \farkasLower{x_i} $ can always be
used to deduce the current tightest bounds $ u'(x_i), l'(x_i) $ from
the ground bounds. A key point is that this holds even if
$u'(x_i)$ and $l'(x_i)$ were discovered by Simplex using previously
tightened bounds for other variables. 

We now prove this claim, by induction, for $ \farkasUpper{x_i} $; the
proof for $ \farkasLower{x_i} $ is symmetrical and is omitted.

\medskip
\textbf{Base case:} 
Suppose no tighter upper bound was learned for the variable $x_i$, so $u'(x_i) $ equals the ground upper bound $u(x_i)$. In this case, $\farkasUpper{x_i}$ is initialized to $\overline{0}$, and $\farkasUpper{x_i}^\intercal \cdot A$ produces a zero row vector. Therefore $c_i=0$, and the upper bound deduced is
\[u'(x_i) = u(x_i).\]
 
Suppose that $u'(x_i)$ is the first bound being
updated dynamically using equation $e$. In this case, for all $j\neq
i$ we have $\farkasLower{x_j}=\farkasUpper{x_j}=\overline{0}$, and so we
get that 
$ \farkasUpper{x_i} = \coef(e) $. Consequently,
\[
  \farkasUpper{x_i}^\intercal \cdot A = e
\]
which indeed produces the equation 
 $ 0 = \sumallvars - x_i $ which was used for the tightening.  We then have $ c_i = -1 $, and
the deduced bound is
\[
  u'(x_i) = \sumposvars{u(x_j)} + \sumnegvars{l(x_j)} + ( -1 + 1 )
  \cdot u_i.
\]

\medskip \textbf{Inductive step:} In the general case,
$ \farkasUpper{x_i} $ is updated using previously tightened
bounds. Consequently, for $j\neq i$, the vectors $\farkasUpper{x_j}$
and $\farkasLower{x_j}$ need not be the zero vectors. In this case we
have that:
\[
  \farkasUpper{x_i}^\intercal \cdot A =
\sumposvars{\farkasUpper{x_j}^{\intercal}} \cdot A +
\sumnegvars{\farkasLower{x_j}^{\intercal}} \cdot A + e
\]

Using the induction hypothesis, we have that each $  \farkasUpper{x_j}^\intercal \cdot A, \farkasLower{x_j}^\intercal \cdot A $
is equal to some vector that represents a row of the form $ 0 = \sum b_{k,j} \cdot x_k $,
where the bound of $ x_j $ is tightened using $ \underset{k \neq j }{\overset{ }{\sum} } b_{k,j} \cdot x_k + ( b_{j,j} + 1 ) \cdot x_j $. 
It is important to note here that it is possible that $ b_{j,i} \neq 0 $,
and thus the coefficient of $ x_i $ in $ \farkasUpper{x_i}^\intercal \cdot A $ 
might not be $ -1 $. 

Based on the above, we get that $ \farkasUpper{x_i}^{\intercal} \cdot A$ produces the equation:

\begin{align*}
  &\textcolor{white}{=\ }
    \sumposvars{ \left[\underset {k \neq j} {\sum} b_{k,j} \cdot x_k +
    b_{j,j} \cdot x_j  \right] } + \\
  &\textcolor{white}{=\ }
  \sumnegvars{ \left[\underset {k \neq j} {\sum } b_{k,j} \cdot
    x_{k,j} + b_{j,j} \cdot x_j \right] } + \\
  &\textcolor{white}{=\ } \sumallvars - x_i \\
  &=
    \sumposvars{ \left[\underset {k \neq j} {\sum} b_{k,j} \cdot x_k +
    ( b_{j,j}+1 ) \cdot x_j \right] } + \\
  &\textcolor{white}{=\ }
    \sumnegvars{ \left[\underset {k \neq j} {\sum} b_{k,j} \cdot x_k  +
  ( b_{j,j} + 1) \cdot x_j \right] } - x_i
  \end{align*}

In other words, the tighter bound that is currently being deduced is
derived using the sum
\begin{align*}
  &\sumposvars{ \left[\underset {k \neq j} {\sum} b_{k,j} \cdot x_k + 
    ( b_{j,j}+1 ) \cdot x_j \right] } + \\
  &\sumnegvars{ \left[\underset {k \neq j} {\sum} b_{k,j} \cdot x_k  +
    ( b_{j,j} + 1) \cdot x_j \right] },
\end{align*}
which, according to the induction hypothesis, has an upper bound of
\[
  \sumposvars{u'(x_j)} + \sumnegvars{l'(x_j)},
\]
exactly as desired --- thus completing the proof.

We end this section with a straightforward algorithm for
extracting $ \coef(e) $ when given $ e $.  We
assume, without loss of generality, that $b$ is the zero vector,
and that the $m$ right-most columns of $A$ form the identity
matrix (if this is initially not so, we add a fresh variable
$x_j$ to every row $j$ of $A$, and set $l(x_j)=u(x_j)=-b_j$.)

\begin{theorem}
  The coefficients of a tableau row $ e $ are equal to its last $m$
  entries, where $m$ is the number of tableau rows.  Formally,
  consider a tableau of the form $ A := [C \mid I] = 0 $, where
  $C \in M_{m \times n-m}(\mathbb{R})$ and $ I $ is the identity
  matrix of size $ m $.  Let $ e = (c,d) $ be a row in the tableau,
  where $d \in \mathbb{R}^m $ is the last $m$ entries of $e$.  Then
  $ d \cdot A = e $
\end{theorem}
\proofname: $ \: $ Upon initialization, $ I $ corresponds to the
coefficients of the last $m$ variables.  Whenever a pivot operation is
applied, the tableau is multiplied by elementary matrices.  Let
$ e = (c, d) $ be a tableau row after several pivots have been
applied, where $ d \in \mathbb{R} ^ m $ corresponds to the last $m$
coefficients of $ e $.  We then have that $ d $
is a row of $ E_k \cdot ... \cdot E_1 \cdot I $ for some sequence of
elementary matrices, or equivalently,
$ d = e_l^\intercal \cdot E_k \cdot ... \cdot E_1 $, and similarly
$ c = e_l^\intercal \cdot E_k \cdot ... \cdot E_1 \cdot C $.  We have
that
$ d \cdot C = e_l^\intercal \cdot E_k \cdot ... \cdot E_1 \cdot C = c
$.
We conclude that $ d \cdot A = d \cdot C \mid d \cdot I = (c,d) = e $. $ \square $ \\

Hence, an efficient algorithm for extracting $ \coef(e) $ from $ e $
is to extract its last $ m $ entries, and transpose that vector.

\section{Proof of Lemma~\ref{thm:maincor}}
\label{app:MainCorProof}

Recall Lemma~\ref{thm:maincor}, which we introduced in Sec.~\ref{sec:SimplexProof}:
\setcounter{theorem}{1}
\begin{lemma}
  \maincor
\end{lemma}
We now give a formal proof for this Lemma.

Since the Simplex calculus that we use is sound and
complete~\cite{KaBaDiJuKo21}, and since it returns a satisfying
assignment $V$ for a satisfiable query, we only need to focus on the
\unsat{} query. Specifically, we need to show the existence of a
vector $w \in \rn{m}$ when the constraints are unsatisfiable.

Observe a row of the tableau, and its corresponding equation $ a =
\underset{j=1}{\overset{n}{\sum}}  c_j \cdot x_j $. The upper bound of
$a$ that can be derived from the equation, denoted $u(a)$, is a linear
combination of the bounds of the $x_j$'s, where we take the upper
bounds for variables with a positive coefficient $c_j$, and the lower
bounds for negative $c_j$ values. More formally,
\[
  u(a) = \underset{j,c_j > 0}{\sum} c_j \cdot u(x_j) + \underset{j,
    c_j < 0}{\sum } c_j \cdot l(x_j)
  \]
The lower bound for $a$ that can be derived from the equation is
marked $l(a)$, and is defined symmetrically.
 
By Lemma~\ref{thm:Lemma1}, the Simplex algorithm concludes that a
query is \unsat{} if and only if it discovers a variable $x_i$ with
inconsistent bounds $l'(x_i) > u'(x_i)$ (whether these are ground
bounds, or bounds discovered through an equation).
Theorem~\ref{thm:mainthm} then states that there exists
$\farkasUpper{x_i},\farkasLower{x_i} \in \rn{m}$ that witness these
bounds.  From the proof of Theorem~\ref{thm:mainthm}, we know that
$\farkasUpper{x_i}^\intercal \cdot A = r$, and we have proved that
$u'(x_i) = u(r + x_i)$.  Similarly, we have that
$\farkasLower{x_i}^\intercal \cdot A = s$ where
$l'(x_i) = l(s + x_i)$.

Consider the vector $ w = \farkasUpper{x_i} - \farkasLower{x_i}$. We
know that $w^\intercal \cdot A = r - s$, and therefore:

\begin{align*}
  u(w^\intercal \cdot A) &= u(r - s ) \\
  &= u(r + x_i - s - x_i) \\
  &\leq  u(r + x_i) - l(s + x_i) \\
  &= u'(x_i) - l'(x_i) < 0
\end{align*}
This means that for any assignment of $V$, the value of
$w^\intercal \cdot A$ is negative.  In other words, we have proved
that for all $ l \leq V \leq u $, it hods that
$ w^\intercal \cdot A \cdot V < 0$, as needed.

\section{Proof production for ReLU bound tightening rules}
\label{app:ReLUProof}

Here, we summarize some bound tightenings
that can be performed as a \relu{} constraint $ f = ReLU(b) $, and
which our method currently supports.  As before, we denote
dynamically-tightened bounds as $ l', u' $, and the ground bounds
as $ l,u $.

As stated before, these tightenings can be introduced as lemmas,
or in some cases be derived as linear tightenings, if appropriate
equations have been added to the tableau (i.e., the equation $f = b$).
Recall that: 
\begin{center}
	$ f = \relu(b) = \begin{cases}
		b & b>0 \\
		0 & \text{otherwise.} \end{cases} $
\end{center}

Therefore, the following tightenings can be performed (provided that
they lead to tighter bounds than the ones presently known):
\begin{center}
	\begin{minipage}{0.8\linewidth}
		\begin{inparaenum}[(i)]
		 	\item For a positive $l'(f)$, $l'(b) := l'(f)$.\\
		 	\item For a positive $l'(b)$, $l'(f) := l'(b)$.\\
		 	\item For any $u'(f)$, $u'(b) := u'(f)$.\\
		 	\item For non-positive $u'(b)$, $u'(f) := 0$. \\ 
		 	\item For a positive $u'(b)$, $u'(f) := u'(b)$.\\
		 \end{inparaenum}
	\end{minipage}
\end{center}

\section{A note on maintaining the correctness of Farkas vectors}
\label{app:MaintainCorrectness}
It may not be immediately clear that updating a ground bound of a
variable $x$ (e.g., as a result of case-splitting) maintains the
correctness of other variables' Farkas vectors that depend on the
ground bounds of $x$. Since the Simplex calculus allows ground bounds
to only be tightened (i.e., lower bounds to be increased, and upper
bounds to be decreased), the dynamically tightened bounds computed
using the ground bounds are tightened as well.  
This can be observed since a dynamic upper bound is calculated using ground upper 
bounds with positive coefficients, and ground lower bounds with negative coefficients. 
Therefore, when ground lower bound increases or ground upper bound decreases,
any dynamic upper bound computed using them decreases too. 
Increasing of a dynamic lower bound follows similarly.
We conclude that using a
Farkas vector to prove a bound, might lead to a tighter bound than
expected, but this does not harm soundness.

For example, consider the case where the Farkas vector of the upper
bound of a variable $ y $ creates the row $ u'(y) = u(x) - l(z) $, and
$ u(x) = 4, \: l(z) = 1 $, so $ u'(y) = 3 $.  If we tighten
$ u(x) := 2 $, then the Farkas vector of the upper bound of $y$ will
prove the bound $ 1 $, which is tighter than $3$.  

\end{document}